\documentclass[prd,preprint,tightenlines,floatfix,showpacs,preprintnumbers,nofootinbib,eqsecnum]{revtex4}
 \usepackage[dvips,final]{graphicx}
  \usepackage{amssymb}

   \usepackage{amsmath}
    \usepackage{amsfonts}

     \usepackage{epsfig}
      \usepackage{bm}

\def\GeV{\,\mbox{GeV}}
\def\mb{\,\mbox{mb}}
\def\TeV{\,\mbox{TeV}}
\def\Pom{{ I\!\!P}}
 \def\Reg{{ I\!\!R}}
 \def\fm{\,\mbox{fm}}

\newcommand\la{\langle}
 \newcommand\ra{\rangle}
 \newcommand\beq{\begin{equation}}
 
 \newcommand\eeq{\end{equation}}
 \newcommand\beqn{\begin{eqnarray}}
 \newcommand\eeqn{\end{eqnarray}}

\begin{document}

\begin{flushright}
LU TP 15-11\\
July 2015
\end{flushright}

\title{Diffractive bremsstrahlung in hadronic collisions}

\author{Roman Pasechnik}
 \email{Roman.Pasechnik@thep.lu.se}

 \affiliation{Department of Astronomy and Theoretical
 Physics, Lund University, SE 223-62 Lund, Sweden}

\author{Boris Kopeliovich}
\email{boris.kopeliovich@usm.cl}

\author{Irina Potashnikova}
\email{irina.potashnikova@usm.cl}

 \affiliation{Departamento de F\'{\i}sica,
Universidad T\'ecnica Federico Santa Mar\'{\i}a; and\\
Centro Cient\'ifico-Tecnol\'ogico de Valpara\'{\i}so,
Avda. Espa\~na 1680, Valpara\'{\i}so, Chile\vspace{1cm}}

\begin{abstract}
\vspace{0.5cm} 
Production of heavy photons (Drell-Yan), gauge bosons, Higgs bosons, heavy flavors, 
which is treated within the QCD parton model as a result of hard parton-parton collision, 
can be considered as a bremsstrahlung process in the target rest frame. In this review, 
we discuss the basic features of the diffractive channels of these processes 
in the framework of color dipole approach. The main observation is a dramatic breakdown 
of diffractive QCD factorisation due to the interplay between soft and hard interactions, 
which dominates these processes. This observation is crucial for phenomenological 
studies of diffractive reactions in high-energy hadronic collisions.
\end{abstract}

\pacs{13.87.Ce, 14.65.Dw, 14.80.Bn}

\maketitle

\section{Introduction}

Diffractive production of particles in hadron-hadron scattering at high energies is one of the basic tools, 
experimental and theoretical, giving access to small-$x$ and nonperturbative QCD physics. 
The characteristic feature of diffractive processes at high energies is the presence of a large rapidity gap 
between the remnants of the beam and target. 

The understanding of the mechanisms of inelastic diffraction came with the pioneering works of Glauber
\cite{Glauber}, Feinberg and Pomeranchuk \cite{FP56}, Good and Walker \cite{GW}. Here diffraction 
is conventionally viewed as a shadow of inelastic processes. If the incoming plane wave contains components
interacting differently with the target, the outgoing wave will have a different composition, i.e. besides 
elastic scattering a new {\it diffractive} state will be created resulting in a new combination of the 
Fock components (for a detailed review on QCD diffraction, see Ref.~\cite{kaidalov-review,KPSdiff}). 
Diffraction, which is usually a soft process, is difficult to predict from the first principles, because it 
involves poorly known nonperturbative effects. Therefore, diffractive reactions characterised by a hard scale
deserve a special attention. It is tempting, on analogy to inclusive reactions, to expect that QCD factorization 
holds for such diffractive processes. Although factorization of short and long distances still holds in diffractive DIS,
the fracture functions are not universal and cannot be used for other diffractive processes.
 
Examples of breakdown of diffractive factorization  are the processes of  production of Drell-Yan dileptons 
\cite{KPST06,Pasechnik:2011nw}, gauge bosons \cite{Pasechnik:2012ac} and heavy flavors \cite{KPST07}.
Factorization turns out to be broken in all these channels in spite of presence of a hard scale given by the 
large masses of produced particles, it occurs due to the interplay of short- and long-range interactions.

The main difficulty in formulation of a theoretical QCD-based framework for diffractive scattering is caused 
by the essential contamination of soft, non-perturbative interactions. For example, diffractive deep-inelastic 
scattering (DIS), $\gamma^*p\to \bar qqp$, although it is a higher twist process, is dominated by soft 
interactions \cite{povh}. Within the dipole approach \cite{zkl} such a process looks like a linear combination 
of elastic scattering amplitudes for $\bar qq$ dipoles of different sizes. Although formally the process 
$\gamma^*\to\bar qq$ is an off-diagonal diffraction, it does not vanish in the limit of unitarity saturation, 
the so called black-disc limit. This happens because the initial and final $\bar qq$ distribution functions are 
not orthogonal. Similar features exhibit the contribution of higher Fock components of the photon, e.g. 
the leading twist diffraction $\gamma^*\to\bar qqg$. 

Diffractive excitation of the beam hadron has been traditionally used as a way to measure the 
Pomeron-hadron total cross section \cite{kaidalov-review}. This idea extended to DIS, allows to measure 
the structure function of the Pomeron \cite{ISh}. The next step, which might look natural, is to assume 
that QCD factorization holds for diffraction, and to employ the extracted parton distributions in the Pomeron 
in order to predict the hard diffraction cross sections in hadronic collisions. However, such predictions for 
hard hadronic diffraction, e.g. high-$p_T$ dijet production, failed by an order of magnitude \cite{tevatron-1,tevatron-2}. 
In this case the situation is different and more complicated, namely, factorization of small and large distances 
in hadronic diffraction is broken because of presence of spectator partons and due to large hadronic sizes.

The cross section of diffractive production of the $W$ boson in $p\bar p$ collisions measured by the CDF 
experiment \cite{cdf1,CDF-WZ}, was also found to be six times smaller than what was predicted relying on 
factorisation and diffractive DIS data \cite{dino}. Besides, the phenomenological models based on diffractive 
factorisation, which are widely discussed in the literature (see e.g. Refs.~\cite{Szczurek,Beatriz}), predict 
a significant increase of the ratio of the diffractive to inclusive gauge bosons production cross sections with energy. 
The diffractive QCD factorisation in hadron collisions is, however, severely broken by the interplay of hard and 
soft dynamics, as was recently advocated in Refs.~\cite{Pasechnik:2011nw, Pasechnik:2012ac}, and this review 
is devoted to study of these important effects within the color dipole phenomenology.

The processes under discussion -- single diffractive Drell-Yan \cite{KPST06,Pasechnik:2011nw}, diffractive radiation 
of vector ($Z,\,W^\pm$) bosons \cite{Pasechnik:2012ac}, diffractive heavy flavor production \cite{KPST07} and 
diffractive associated heavy flavor and Higgs boson production \cite{Pasechnik:2014lga} -- correspond to off-diagonal 
diffraction. While diagonal diffraction is enhanced by absorption effects (in fact it is a result of absorption), 
the off-diagonal diffractive processes are suppressed by absorption, and even vanish in the limit of maximal 
absorption, i.e. in the black-disc limit.

The absorptive corrections, also known as the survival probability of rapidity gaps \cite{enh-2}, are related to 
initial- and final-state interactions. Usually the survival probability is introduced into the diffractive cross section 
in a probabilistic way \cite{Ryskin:2011qe} and is estimated in simplified models such as eikonal, quasi-eikonal, 
two-channel approximations, etc. 

According to the Good-Walker basic mechanism of diffraction, the off-diagonal diffractive amplitude is a linear 
combination of diagonal (elastic) diffractive amplitudes of different Fock components in the projectile hadron.
Thus, the absorptive corrections naturally emerge at the amplitude level as a result of mutual cancellations 
between different elastic amplitudes. Therefore, there is no need to introduce any additional multiplicative gap 
survival probability factors. Within the light-cone color dipole approach \cite{zkl} a diffractive process 
is considered as a result of elastic scattering of $\bar qq$ dipoles of different sizes emerging in incident 
Fock states. The study of the diffractive Drell-Yan reaction performed in Ref.~\cite{KPST06} has revealed 
importance of soft interactions with the partons spectators, which contributes on the same footing as hard 
perturbative ones, and strongly violate QCD factorization.

One of the advantages of the dipole description is the possibility to calculate directly (although in a process-dependent way) 
the full diffractive amplitude, which contains all the absorption corrections by employing the phenomenological universal 
dipole cross section (or dipole elastic amplitudes) fitted to DIS data. The gap survival amplitude can be explicitly singled out as 
a factor from the diffractive amplitude being a superposition of dipole scatterings at different transverse separations.

Interesting, that besides interaction with the spectator projectile partons, there is another important source for diffractive 
factorization breaking. Even a single quark, having no spectator co-movers, cannot radiate Abelian fields ($\gamma,\,Z,\,W^\pm,H$) 
interacting diffractively with the target with zero transverse momentum transfer \cite{KST-par}, i.e. in forward direction scattering.
This is certainly contradicts the expectations based of diffractive factorization. In the case of a hadron beam the forward directions 
for the hadron and quark do not coincide, so a forward radiation is possible, but is strongly suppressed (see below).

Interaction with the spectator partons opens new possibilities for diffractive radiation in forward direction, namely the transverse 
momenta transferred to different partons can compensate each other.  It was found in Refs.~\cite{KPST06,Pasechnik:2011nw,Pasechnik:2012ac} 
that this contribution dominates the forward diffractive Abelian radiation cross section. This mechanism leads to a dramatic violation 
of diffractive QCD factorisation, which predicts diffraction to be a higher twist effect, while it turns out to be a leading twist effect 
due to the interplay between the soft and hard interactions.  Although diffractive gluon radiation off a forward quark does not vanish
due to possibility of glue-glue interaction, the diffractive factorisation breaking in non-Abelian radiation is still important.

In this review, we briefly discuss the corresponding effects whereas more details can be found in Refs.~\cite{KPST06,KPST07,Pasechnik:2011nw,
Pasechnik:2012ac,Pasechnik:2014lga}.

\section{Color dipole picture of diffractive excitation}

Single diffractive scattering and production of a new (diffractive) state, i.e. diffractive excitation, 
emerges as a consequence of quantum fluctuations in projectile hadron. The orthogonal hadron
state $|h\rangle$ can be excited due to interactions but can be decomposed over 
the orthogonal and complet set of eigenstates of interactions $|\alpha\rangle$ 
as \cite{Kopeliovich:1978qz,Miettinen:1978jb,zkl}
\beqn
|h\rangle = \sum C^h_\alpha\,|\alpha\rangle \,, \qquad \hat f_{el}|\alpha\rangle=f_\alpha |\alpha\rangle\,,
\eeqn
where $\hat f_{el}$ is the elastic amplitude operator and $f_{\alpha}$ is one of its eignestates. 
The eigen amplitudes $f_\alpha$ are the same for different types of hadrons. Hence, the elastic 
$h\to h$ and single diffractive $h\to h'$ amplitudes can be conveniently written in terms of the
elastic eigen amplitudes $f_\alpha$ and coefficients $C_\alpha^h$, i.e.
\beqn
f^{hh}_{el}=\sum|C_\alpha^h|^2\,f_\alpha\,, \qquad f^{hh'}_{sd}=\sum (C^{h'}_\alpha)^*C^h_\alpha\, f_\alpha \,,
\eeqn
respectively, such that the forward single diffractive cross section
\beqn
\sum_{h'\not=h}\frac{d\sigma_{sd}}{dt}\Big|_{t=0}=
\frac{1}{4\pi}\Big[\sum_{h'}|f_{sd}^{hh'}|^2-|f_{el}^{hh}|^2\Big]=
\frac{\langle f_\alpha^2 \rangle-\langle f_\alpha \rangle^2}{4\pi}
\eeqn
is given by the dispersion of the eigenvalues distribution. 

It was suggested in Ref.~\cite{zkl} that eigenstates of QCD interactions 
are color dipoles, such that any diffractive amplitude can be considered as
a superposition of universal elastic dipole amplitudes. Such dipoles experience
only elastic scattering and characterized only by transverse separation $\vec r$.
The total hadron-proton cross section is then given by its eigenvalue, 
the universal dipole cross section  
\beqn
\sigma(\vec r)\equiv \int d^2b\,2{\rm Im} f_{el}(\vec b,\vec r)\,,
\eeqn
as follows
\beqn
\sigma_{tot}^{hp}=\sum |C_\alpha^h|^2\sigma_\alpha = \int d^2r |\Psi_h(\vec r)|^2\sigma(\vec r)\equiv
\langle\sigma(\vec r)\rangle \,,
\eeqn
where $\Psi_h(\vec r)$ is the ``hadron-to-dipole'' transition wave function (incident parton momentum fractions are omitted).
The dipole description of diffraction is based on the fact that dipoles of different
transverse size $r_{\perp}$ interact with different cross sections
$\sigma(r_{\perp})$, leading to the single inelastic diffractive
scattering with a cross section, which in the forward limit is given
by \cite{zkl},
\begin{eqnarray}
\frac{\sigma_{sd}}{dp_{\perp}^2}\Bigg|_{p_{\perp}=0}=
\frac{\langle\sigma^2(\vec r)\rangle-\langle\sigma(\vec r)\rangle^2}{16\pi},\,
\end{eqnarray}
where $p_{\perp}$ is the transverse momentum of the recoil proton,
$\sigma(r)$ is the universal dipole-proton cross section, and 
operation $\langle\dots\rangle$ means averaging over
the dipole separation. For low and moderate energies, $\sigma(r)$ 
also depends on Bjorken variable $x$ whereas
in the high energy limit, the collision c.m. energy squared $s$ is a more appropriate variable \cite{KST-par,GBWdip}.
The phenomenological dipole cross section fitted to data on inclusive DIS implicitly incorporates the effect
of gluon bremsstrahlung. The latter is more important on a hard scale, this is why the small-distance 
dipole cross section rises faster with $1/x$.
\begin{figure}[h!]
\centerline{\epsfig{file=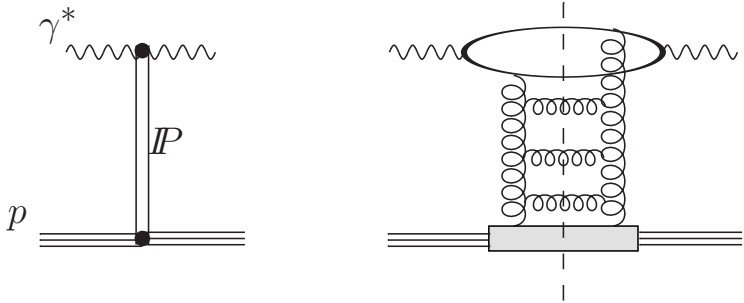,width=10cm}} 
\caption{The DIS cross section via phenomenological Pomeron exchange (left) and 
a perturbative QCD ladder (right). At small $x$ the virtual photon fluctuates 
into a $q\bar q$ dipole and more complicated Fock states which then interact 
with the hadronic target.}
\label{fig:DIS}
\end{figure}

Even for the simplest quark-anitiquark dipole configuration, a theoretical prediction of 
the partial dipole amplitude  $f^{q\bar q}_{el}(\vec b,\vec r)$ and the dipole cross section $\sigma_{q\bar q}(\vec r)$
from the first QCD principles is still a big challenge so these are rather fitted to data. The universality of 
the dipole scattering, however, enables us to fit known parameterizations to one set of known data 
(e.g. inclusive DIS) and use them for accurate predictions of other yet unknown observables (e.g. rapidity gap 
processes). 
\begin{figure}[h!]
\centerline{\epsfig{file=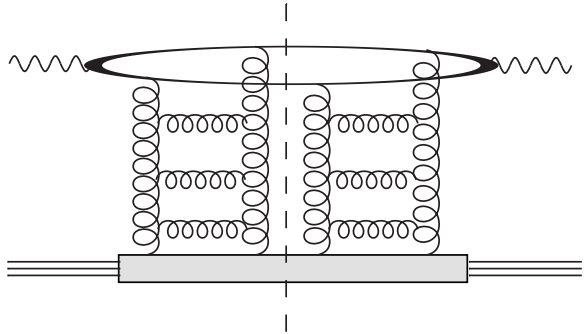,width=8cm}} 
\caption{The diffractive DIS cross section via double 
ladder exchange.}
\label{fig:DDIS}
\end{figure}

Indeed, at small Bjorken $x$ in DIS the virtual photon exhibits partonic structure as shown
in Fig.~\ref{fig:DIS}. The leading order configuration, the $q\bar q$ dipole, then elastically rescatters off 
the proton target $p$ providing a phenomenological access to $\sigma_{q\bar q}(\vec r,x)$. When it comes to
diffractive DIS schematically represented in Fig.~\ref{fig:DDIS}, the corresponding single diffractive 
cross section in the forward proton limit $t\to 0$ is given by the dipole cross section squared, i.e.
\beqn
16\pi\, \frac{d\sigma_{sd}^{\gamma^*p}(x,Q^2)}{dt}\Big|_{t=0}=
\int d^2 r \int_0^1 d\alpha\, |\Psi_{\gamma^*}(\vec r,\alpha,Q^2)|\,\sigma_{q\bar q}^2(\vec r,x)\,,
\eeqn
where $\alpha$ is the light-cone momentum fraction of the virtual photon carried by the quark. Here, 
the dipole size $\vec r$ is regulated by the photon light-cone wave function $\Psi_{\gamma^*}$ which
can be found e.g. in Ref.~\cite{Bjorken:1970ah}. The mean dipole size squared is inversely proportional
to the quark energy squared
\beqn
\langle r^2 \rangle \sim \frac{1}{\epsilon^2}=\frac{1}{Q^2\alpha(1-\alpha)+m_q^2} \,.
\eeqn
The dipole size is assumed to be preserved during scattering in the high energy limit.

Hard and soft hadronic fluctuations have small $\langle r^2 \rangle\sim 1/Q^2$ 
(nearly symmetric $\alpha\gg m_q^2/Q^2$ configuration) and large 
$\langle r^2 \rangle\sim 1/m_q^2$, $m_q\sim \Lambda_{\rm QCD}$ (aligned jet $\alpha\sim m_q^2/Q^2$ 
configuration) sizes, respectively. Remarkably enough, soft fluctuations play a dominant role 
in diffractive DIS in variance with inclusive DIS \cite{povh}. 
Although such soft fluctuations are very rare, their interactions with the target occur with a large cross section
$\sigma \sim 1/m_q^2$ which largely compensate their small $\sim m_q^2/Q^2$ weights. 
On the other hand, abundant hard fluctuations with nearly 
symmetric small-size dipoles $\langle r^2 \rangle\sim 1/Q^2$ have vanishing (as $1/Q^2$) cross section.
It turns out that in inclusive DIS, both hard and soft contributions to the total cross section behave
as $1/Q^2$ (semi-hard and semi-soft), while in diffractive DIS the soft fluctuations $\sim 1/m_q^2Q^2$
dominate over the hard ones $\sim 1/Q^4$. This also explains why the ratio $\sigma_{sd}/\sigma_{inc}$ in DIS
is nearly $Q^2$ independent as well as a higher-twist nature of the diffractive DIS.

The main ingredient of the dipole approach is the phenomenological dipole cross section, 
which is parameterized in the saturated form \cite{GBWdip},
\begin{eqnarray}
\sigma_{\bar qq}(r,x)=\sigma_0(1-e^{-r_p^2/R_0^2(x)})\,,
\label{fel}
\end{eqnarray}
and fitted to DIS data. Here, $x$ is the Bjorken variable, $\sigma_0=23.03\mb$ and 
$R_0(x)=0.4\fm\times(x/x_0)^{0.144}$, where $x_0=0.003$. In $pp$ collisions $x$ is identified with gluon $x_2=M^2/x_1 s \ll 1$ where $M$ is the invariant 
mass of the produced system and $s$ is the $pp$ c.m. energy. This simplified parametrization (cf. Ref.~\cite{bartels}), 
appeared to be quite successful providing a reasonable description of HERA (DIS and DDIS) data. 

In soft processes, however, the Bjorken variable $x$ makes no sense, and gluon-target collision c.m. 
energy squared $\hat s=x_1 s$ ($s$ is the $pp$ c.m. energy) 
is a more appropriate variable, while the saturated form (\ref{fel}) should be retained \cite{KST-par}. 
The corresponding parameterisations for $\sigma_0=\sigma_0(\hat s)$ and $R_0=R_0(\hat s)$ read
\begin{eqnarray}\nonumber
 R_0(\hat s)=0.88\,\mathrm{fm}\,(s_0/\hat s)^{0.14}\,,\quad
 \sigma_0(\hat s)=\sigma_{tot}^{\pi p}(\hat s)
 \Big(1+\frac{3R_0^2(\hat s)}{8\langle r_{ch}^2 \rangle_{\pi}}\Big)\,.
 \label{KST-params}
\end{eqnarray}
where the pion-proton total cross section is parametrized as
\cite{barnett} $\sigma_{tot}^{\pi p}(\hat s)=23.6(\hat s/s_0)^{0.08}$ mb,
$s_0=1000\,\GeV^2$, the mean pion radius squared is \cite{amendolia}
$\langle r_{ch}^2 \rangle_{\pi}=0.44$ fm$^2$. An explicit analytic form 
of the $x$- and $\hat s$-dependent parameterisations for the elastic amplitude 
$f_{el}(\vec{b},\vec{r})$ accounting for an information about the dipole 
orientation w.r.t. the color background field (i.e. the angular dependence 
between $\vec r$ and $\vec b$) can be found in Refs.~\cite{GBW-par,kpss,KST-GBW-eqs}.

The ansatz (\ref{fel}) incorporate such important phenomenon as saturation 
at a soft scale since it levels off at $r\gg R_0$. Another important 
feature is vanishing of the cross section at small $r\to 0$ as 
$\sigma_{q\bar q}\propto r^2$ \cite{zkl}. This is a general 
property called color transparency which reflects the fact that a point-like colorless 
object does not interact with external color fields. Finally, the quadratic
$r$-dependence is an immediate consequence of gauge invariance and
nonabeliance of interactions in QCD.

\section{Diffractive Abelian radiation: Regge vs dipole approach}

\subsection{Diffractive factorisation}

The cross section of the inclusive Drell-Yan (DY) is expressed via the dipole cross section
in a way similar to DIS \cite{KRT00}
\beqn
\frac{d\sigma_{\rm DY}(qp\to \gamma^*X)}{d\alpha dM^2}=\int d^2r 
|\Psi_{q\gamma^*}(\vec r,\alpha)|^2\,\sigma_{q\bar q}(\alpha\vec r)\,,
\eeqn
where $\alpha$ is the light-cone momentum fraction carried by the heavy 
photon off the parent quark. QCD factorisation relates inclusive DIS with 
DY, and similarity between these processes is the source of universality of the
hadron PDFs.
\begin{figure*}[!h]
\begin{minipage}{1.0\textwidth}
 \centerline{\includegraphics[width=0.75\textwidth]{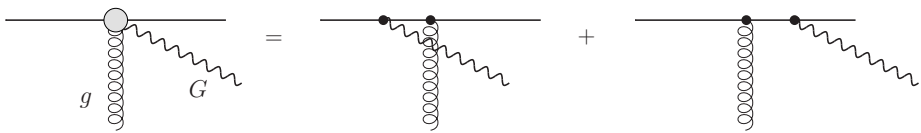}}
\end{minipage}
   \caption{Gauge boson radiation by a projectile quark in the target rest frame.}
 \label{fig:vertex}
\end{figure*}

Now, consider the forward single diffractive Drell-Yan (DDY) and vector 
bosons production $G=Z,\,W^\pm$ in $pp$ collisions
which is characterized by a relatively small momentum transfer between 
the colliding protons. In particular, one of the protons, e.g. $p_1$, radiates a hard 
virtual gauge $G^*$ boson with $k^2=M^2\gg m_p^2$ and hadronizes into a hadronic
system $X$ both moving in forward direction and separated by a large 
rapidity gap from the second proton $p_2$, which remains intact. In the DDY case,
\begin{eqnarray}
p_1+p_2\to X+(gap)+p_2 \,, \qquad X\equiv\gamma^*(l^+l^-)+Y \,.
\end{eqnarray}
Both the di-lepton and $X$, the debris of $p_1$, stay in the forward
fragmentation region. In this case, the virtual photon is predominantly 
emitted by the valence quarks of the proton $p_1$.

In some of the previous studies \cite{Szczurek,Landshoff-DDY} of 
the single diffractive Drell-Yan reaction the analysis was made 
within the phenomenological Pomeron-Pomeron and $\gamma$-Pomeron 
fusion mechanisms using the Ingelman-Shlein approach \cite{ISh} 
based on diffractive factorization. In analogy to the proven collinear 
factorisation \cite{Collins:1989gx} for inclusive processes, one assumes factorization of 
short and long distances in diffractive processes characterized by a hard scale. Besides one assumes 
that the soft part of the interaction, is carried out by Pomeron exchange, which is universal for different diffractive processes,
i.e. Regge factorization is assumed as well. That could be true if the Pomeron were a true Regge pole, what is not supported by 
any known dynamical model. The above two assumptions lead to the following form of the diffractive DY cross section 
\cite{Landshoff-DDY,Donnachie:1987gu} expresses in terms of the Pomeron PDFs $F_{\bar q/I\!\!P}$
\beqn \label{sd-DY}
\sigma^{\rm DY}_{sd}=G_{I\!\!P/p}\otimes F_{\bar q/I\!\!P}\otimes 
F_{q/p}\otimes \hat{\sigma}(q\bar q\to \bar ll) \,.
\eeqn
The diffractive factorisation leads to specific features 
of the differential DY cross sections similar to those in diffractive DIS 
process, e.g., a slow increase of the diffractive-to-inclusive DY cross 
sections ratio with c.m.s. energy $\sqrt{s}$, its practical independence 
on the hard scale, the invariant mass of the lepton pair squared, 
$M^2$ \cite{Szczurek}.

However, presence of spectator partons in hadronic collisions
leads to a dramatic breakdown of diffractive factorization of short 
and long distances. On the contrary to inclusive processes, where spectator 
partons do not participate in the hard reactions in leading order, below we demonstrate that
in diffraction the spectator partons do participate on a soft scale, i.e. their contribution is 
enhanced by $Q^2/\Lambda^2$. In particular, the spectator partons generate large 
absorptive corrections, usually called rapidity gap survival probability, which cause 
a strong suppression of the diffractive cross section compared with Eq.~(\ref{sd-DY}).
\begin{figure}[h!]
\centerline{\epsfig{file=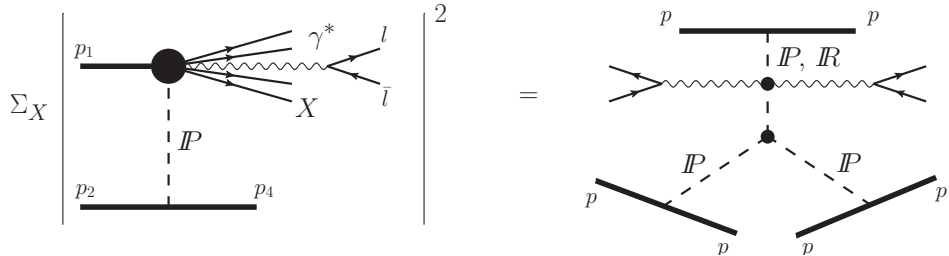,width=13cm}} \caption{The
diffractive DY cross section summed over
excitation channels at fixed effective mass $M_X$ (left panel). The latter
corresponds to the Mueller graph in Regge picture (right panel).}
\label{fig:Pom}
\end{figure}

One can derive a Regge behavior of the diffractive cross section
of heavy photon production in terms of the usual light-cone variables,
\begin{eqnarray}
x_{1} = \frac{p_{\gamma}^+}{p_1^+};\ \ \ \ \
x_{2} = \frac{p_{\gamma}^-}{p_2^-},
\label{x12}
\end{eqnarray}
so that $x_{1}x_{2}=(M^2+k_T^2)/s$ and $x_{1}-
x_{2}=x_{F}$, where $M$, $k_T$ and $x_{F}$ are
the invariant mass, transverse momentum and Feynman variable of the
heavy photon (di-lepton). 

\begin{figure*}[!h]
\begin{minipage}{0.8\textwidth}
 \centerline{\includegraphics[width=1.0\textwidth]{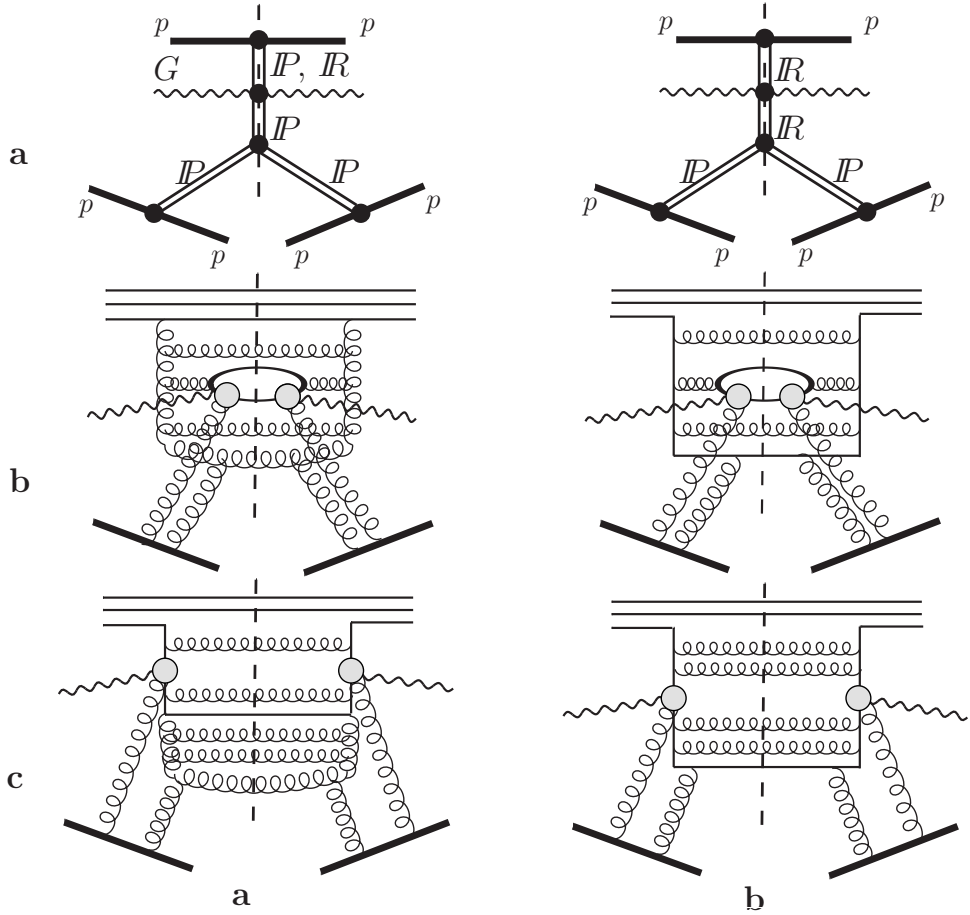}}
\end{minipage}
   \caption{
\small The upper row contains the triple-Regge graphs for
$pp\to (XG^*)+p$. A few key examples of diagrams for diffractive
excitation of a large invariant mass are given by 2d and 3rd rows.}
 \label{fig:3-regge}
\end{figure*}

In the limit of small $x_{1}\to0$ and
large $z_p\equiv p_4^+/p_2^+\to 1$  the diffractive DY cross section
is given by the Mueller graph shown in Fig.~\ref{fig:Pom}. In this
case, the end-point behavior is dictated by the following general
result
\begin{eqnarray}
\frac{d\sigma}{dz_p dx_{1} dt}\Big|_{t\to0}\propto
\frac{1}{(1-z_p)^{2\alpha_{\Pom}(t)-1}x_{1}^{\varepsilon}}\,,
\end{eqnarray}
where $\alpha_{\Pom}(t)$ is the Pomeron trajectory corresponding to
the $t$-channel exchange, and $\varepsilon$ is equal to 1 or 1/2 for
the Pomeron $\Pom$ or Reggeon $\Reg$ exchange corresponding to
$\gamma^*$ emission from sea or valence quarks, respectively.
Thus, the diffractive Abelian radiation process $pp\to (X\to G^*+Y)p$ 
at large Feynman $x_F\to1$, or small 
\beq 
 \xi=1-x_F=\frac{M_X^2}{s}\ll1, 
 \label{xi} 
\eeq
 is described by triple Regge graphs in Fig.~\ref{fig:3-regge} where 
 we also explicitly included radiation of a virtual gauge boson $G^*$.
The Feynman graphs corresponding to the corresponding triple-Regge
terms, are shown in Fig.~\ref{fig:3-regge} (second and third rows). 
The (ba) and (ca) diagrams illustrate the
3-Pomeron term, i.e. 
 \beq
 \frac{d\sigma_{diff}^{\Pom\Pom\Pom}} {d\xi dt} 
 \propto  \xi^{-\alpha_{\Pom}(0)-2\alpha^\prime_{\Pom}(t)} \,.
 \label{PPP} 
 \eeq
It is worth to mention that the smallness of the triple-Pomeron coupling 
is related to the known shortness of gluon correlation length. 
The amplitude $q+g\to q+G$ is given by open circles as in Fig.~\ref{fig:vertex}.
So the 3-Pomeron term is interpreted as an excitation of the projectile proton 
due to the gluon radiation. The diffractive valence quark excitation is shown 
in Fig.~\ref{fig:3-regge}, graphs (bb) and (cb) and contributes to 
 \beq
 \frac{d\sigma_{diff}^{\Pom\Pom\Reg}}{d\xi dt} \propto 
 \xi^{\alpha_{\Reg} (0)-\alpha_{\Pom} (0)-2\alpha^\prime_{\Pom}(t)} \,. \label{PPR} 
 \eeq 

\subsection{Diffractive factorisation breaking in forward diffraction}

As an alternative to the diffractive factorization based approach, the dipole
description of the QCD diffraction, was presented in Refs.~\cite{zkl} (see also
Ref.~\cite{BBGG81}). The color dipole description of inclusive Drell-Yan process was first 
introduced in Ref.~\cite{deriv1} (see also Refs.~\cite{BHQ,KRT00}) and treats 
the production of a heavy virtual photon via Bremsstrahlung mechanism 
rather than $\bar qq$ annihilation. The dipole approach applied to 
diffractive DY reaction in Refs.~\cite{KPST06,Pasechnik:2011nw} and later 
in diffractive vector boson production \cite{Pasechnik:2012ac} has explictly 
demonstrated the diffractive factorisation breaking in diffractive Abelian 
radiation reactions. 

It is worth emphasizing that the quark radiating the gauge boson
cannot be a spectator, but must participate in the interaction. This
is a straightforward consequence of the Good-Walker mechanism of
diffraction \cite{GW}. According to this picture, diffraction vanishes if all
Fock components of the hadron interact with the same elastic amplitudes.
Then an unchanged Fock state composition emerges from the interaction,
i.e. the outgoing hadron is the same as the incoming one, so the interaction
is elastic. 
\begin{figure}[h!]
\centerline{\epsfig{file=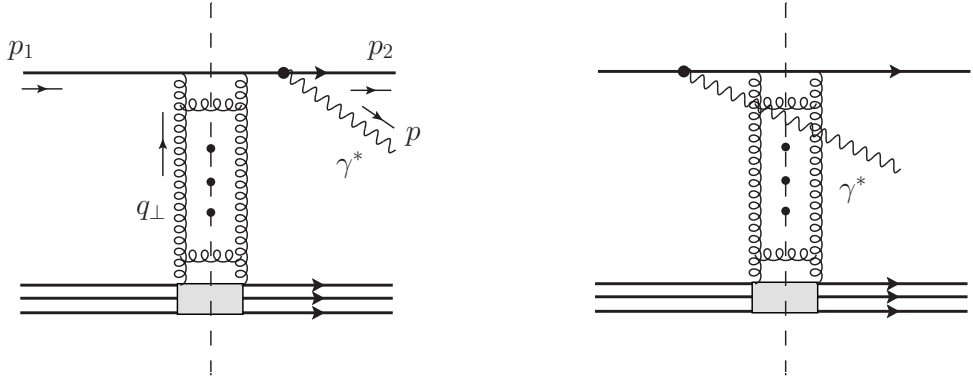,width=13cm}} \caption{Schematic illustration of the
typical contributions and kinematics of the diffractive Drell-Yan process in the quark-target
collision.} \label{fig:gam-kin}
\end{figure}

For illustration, consider diffractive photon radiation off a quark \cite{KST-par}. The relevant contributions 
and kinematics of the process are schematically presented in Fig.~\ref{fig:gam-kin} where the Pomeron
exchange is depicted as an effective two-gluon (BFKL) ladder.
The corresponding framework has previously been used for
diffractive gluon radiation and diffractive DIS processes 
in Refs.~\cite{KST-par,Raufeisen:2002zp,kst1} and we adopt 
similar notations in what follows. Applying the generalized optical theorem in the high energy 
limit with a cut between the ``screening'' and ``active'' gluon as shown by dashed line in Fig.~\ref{fig:gam-kin} 
we get,
\begin{eqnarray} \nonumber
\hat{A}_{SD}&=&\frac{i}{2} \sum_{Y_8^*} \Big[ \hat{A}^\dagger(q\gamma p\to q\{Y_8^*\})
\hat{A}(qp\to q\{Y_8^*\}) \\
&+&  \hat{A}^\dagger(q\gamma p\to q\gamma\{Y_8^*\})
\hat{A}(qp\to q\gamma\{Y_8^*\}) \Big] \,,
\label{D-amp}
\end{eqnarray}
with summation going through all octet-changed intermediate states $\{Y_8^*\}$.
In Eq.~(\ref{D-amp}), the first and second terms correspond to the first and second
diagrams in Fig.~\ref{fig:gam-kin}.
Then we switch to impact parameter representation,
\beq
 \hat{A}(\vec b,\vec r) = \frac{1}{(2\pi)^4}\int d^2\vec q_\perp\,d^2\vec \kappa\;
 \hat{A}(\vec q_\perp,\vec\kappa)\,  e^{ - i\vec q_\perp\cdot\vec b - i\vec \kappa\cdot\vec r }\,, \qquad \vec\kappa = \alpha \vec{p}_2 - (1-\alpha)\vec{p} \,, 
 \eeq
where $\vec q_\perp,\,\vec p_2,\, \vec p$ are the transverse momenta of the Pomeron, final quark and the radiated photon $\gamma^*$,
$\alpha$ is the longitudinal momentum fraction of the photon taken off the parent quark $p_1$, and $\kappa$ is the relative transverse momentum
between the final quark and $\gamma^*$.
Thus, the amplitude of the ``screening'' gluon exchange summed over projectile 
valence quarks $j=1,2,3$ reads
\beqn \nonumber
&&\hat{A}(qp\to q\{Y_8^*\})=\sum_{a} \tau_a\langle f|\hat{\gamma}_a(\vec b_1)| i \rangle \,, \quad
\hat{A}(q\gamma p\to q\gamma\{Y_8^*\})=\sum_{a} \tau_a\langle f|\hat{\gamma}_a(\vec b_2)| i \rangle \,, \\
&&\hat{A}(qp\to q\gamma\{Y_8^*\})=\hat{A}(q\gamma p\to q\gamma\{Y_8^*\})=
\sum_{a} \tau_a\Big[\langle f|\hat{\gamma}_a(\vec b_1)| i \rangle - 
\langle f|\hat{\gamma}_a(\vec b_2)| i \rangle\Big]\, \Psi_{q\to q\gamma}(\vec r,\alpha) \,, \quad
\nonumber
\eeqn
where $\vec b_1\equiv \vec b$ and $b_2\equiv \vec b - \alpha\vec r$ are 
the impact parameter of the quark before and after photon radiation, $\vec r$ is the
transverse separation between the quark and the radiated photon, $\alpha$ is the
momentum fraction taken by the photon, $\Psi_{q\to q\gamma}$ is the distribution
function for the $q\gamma$ fluctuation of the quark,
$\lambda_a=2\tau_a$ are the Gell-Mann matrices from a gluon coupling to the quark, 
and the matrices $\hat{\gamma}_a$
are the operators in coordinate and color space for the target quarks,
\[
\hat{\gamma}_a(\vec R)=\sum_i \tau^{(i)}_a \chi(\vec R-\vec s_i)\,, \qquad 
\chi(\vec s)\equiv \frac{1}{\pi} \int d^2q \frac{\alpha_s(q) e^{i\vec q\cdot\vec s}}{q^2+\Lambda^2},
\]
which depend on the effective gluon mass $\Lambda\sim 100$ MeV, and on
the transverse distance between $i$-th valence quark in the target nucleon 
and its center of gravity, $\vec s_i$. 

Combining these ingredients into the diffractive amplitude (\ref{D-amp})
one should average over color indices of the valence quarks and
their relative coordinates in the target nucleon $|3q\rangle_1$. The color
averaging results in,
\[
\langle \tau_a^{(j)}\cdot \tau_{a'}^{(j')} \rangle_{|3q\rangle_1} = 
\left\{
     \begin{array}{lr}
       \frac{1}{6}\delta_{aa'} & : j=j'\\
       -\frac{1}{12}\delta_{aa'} & : j\not=j'.
     \end{array}
   \right.
\]
Finally, averaging over quark relative coordinates $\vec s_i$ leads to
\[
\langle i|\hat{\gamma}_a(\vec b_k)\hat{\gamma}_{a'}(\vec b_l)| i 
\rangle_{|3q\rangle_1} =\frac{3}{4} \delta_{aa'} S(\vec b_k,\vec b_l)\,,
\]
where $S(\vec b_k,\vec b_l)$ is a scalar function, which can be expressed in terms of 
the quark-target scattering amplitude $\chi(\vec r)$ and the proton wave function \cite{KST-par}. 
Then, the total amplitude,
\[
\hat{A}(\vec b,\vec r)\propto S(\vec b, \vec b) - S(\vec b - \alpha \vec r, \vec b - \alpha \vec r) \,.
\]
After Fourier transform one notices that in the forward quark limit $q_\perp\to 0$ the amplitude for single 
diffractive photon or any Abelian radiation vanishes, $A(\vec q_\perp,\kappa)|_{q_\perp\to0}=0$, and
\[
\frac{d\sigma^{\rm DY}_{sd}}{d\alpha dq_T^2}\Big|_{q_T=0}=0\,,
\]
in accordance with the Landau-Pomeranchuk principle. Indeed, in both Fock components 
of the quark $|q\rangle$ and $|q\gamma^*\rangle$, only the quark interacts, so these 
components interact equally and thus no diffraction is possible. One immediately concludes 
that the diffractive factorisation must be strongly broken.

The function $S(\vec b_k,\vec b_l)$ above is directly related to the $q\bar q$ dipole 
cross section as,
\beq
\sigma_{\bar q q}(\vec{r}_1-\vec{r}_2)\equiv \int d^2 b \Big[ S(\vec b+\vec{r}_1,\vec b+\vec{r}_1)+
S(\vec b+\vec{r}_2,\vec b+\vec{r}_2)-2S(\vec b+\vec{r}_1,\vec b+\vec{r}_2)  \Big]\,.
\eeq
Thus, following anological scheme one can obtain the diffractive amplitude of any diffractive process 
as a linear combination of the dipole cross sections for different dipole separations.
As was anticipated, the diffractive amplitude represents the destructive interference effect from 
scattering of dipoles of slightly different sizes. Such an interference results in an interplay 
between hard and soft fluctuations in the diffractive $pp$ amplitude, contributing to 
breakdown of diffractive factorisation.

When one considers diffractive DY off a finite-size object like a proton, in both 
Fock components, $|3q\rangle$ and $|3q\gamma^*\rangle$, only the quark hadron-scale dipoles
interact. These dipoles are large due to soft intrinsic motion of quarks in the
projectile proton wave function. The dipoles, however, have different sizes,
since the recoil quark gets a shift in impact parameters. So the dipoles interact
differently giving rise to forward diffraction. The contribution of a
given projectile Fock state to the diffraction amplitude is given by
the difference of elastic amplitudes for the Fock states including
and excluding the gauge boson, 
\beq 
 \Im f^{(n)}_{diff}=\Im f^{(n+G)}_{el}-\Im f^{(n)}_{el}\,, 
 \label{f-diff}
\eeq 
where $n$ is the total number of partons in the Fock state; $f^{(n+G)}_{el}$ and
$f^{(n)}_{el}$ are the elastic scattering amplitudes for the whole
$n$-parton ensemble, which either contains the gauge boson or does
not, respectively. Although the gauge boson does not participate in
the interaction, the impact parameter of the quark radiating the
boson gets shifted, and this is the only reason why the difference
Eq.~(\ref{f-diff}) is not zero (see the next section). This also conveys that this quark
must interact in order to retain the diffractive amplitude nonzero
\cite{KPST06,Pasechnik:2011nw}. For this reason in the graphs depicted in
Fig.~\ref{fig:3-regge}  the quark radiating $G$ always takes part in
the interaction with the target.

Notice that there is no one-to-one correspondence between
diffraction in QCD and the triple-Regge phenomenology. In
particular, there is no triple-Pomeron vertex localized in rapidity.
The colorless ``Pomeron'' contains at least two $t$-channel gluons,
which can couple to any pair of projectile partons. For instance in
diffractive gluon radiation, which is the lowest order term in the
triple-Pomeron graph, one of the $t$-channel gluons can couple to
the radiated gluon, while another one couples to another parton at
any rapidity, e.g. to a valence quark (see Fig.~3 in
\cite{KST-par}). Apparently, such a contribution cannot be
associated literally with either of the Regge graphs in
Fig.~\ref{fig:3-regge}. Nevertheless, this does not affect much the
$x_F$- and energy dependencies provided by the triple-Regge graphs,
because the gluon has spin one.

It is also worth mentioning that in Fig.~\ref{fig:3-regge} we
presented only the lowest order  graphs with two gluon exchange. The
spectator partons in a multi-parton Fock component also can interact
and contribute to the elastic amplitude of the whole parton
ensemble. This gives rise to higher order terms, not shown
explicitly in Fig.~\ref{fig:3-regge}. They contribute to the
diffractive amplitude Eq.~(\ref{f-diff}) as a factor, which we
define as the gap survival amplitude \cite{Pasechnik:2012ac}.

As was mentioned above the diffractive Abelian radiation off a
dipole is non-vanishing in the forward domain which is different, for instance, 
from the $q\to q + \gamma^*$ case (see e.g. Refs.~\cite{KST-par, KPST06}). 
Indeed, it is well-known that the off-diagonal diffraction is realised 
as long as different Fock states in the projectile hadron 
have different elastic interaction amplitudes \cite{Glauber, FP56, GW,
KPSdiff}. Due to the fluctuation
$|q\rangle \to |qG\rangle$ the quark changes its position in the
transverse plane by $\Delta\vec{r}=\alpha\vec{r}$ where $\vec{r}$ 
is the quark-boson transverse separation. Integrating over the impact 
parameter one realises that the Fock states corresponding to a single 
quark and a quark plus a boson interact with the same cross section
such that a quark does not radiate at zeroth transverse momentum
transfer. This happens under the assumption that the coherence
time with respect to the radiation is much larger than $\Delta t$ scale
between the subsequent interactions valid at forward rapidities.
This is the main (model-independent) reason why diffractive 
production of $G$ in the forward direction never realises (for more
details, see Ref. \cite{KST-par, BH}).

The disappearance of both inelastic 
and diffractive forward Abelian radiation has a direct analogy in QED:
if the electric charge gets no ``kick'', i.e. is not accelerated, no 
photon is radiated, provided that the radiation time considerably 
exceeds the duration time of interaction. This is dictated by the 
renown Landau-Pomeranchuk principle \cite{LP}: radiation depends 
on the strength of the accumulated kick, rather than on its structure, 
if the time scale of the kick is shorter than the radiation time.
It is worth to notice that the non-Abelian QCD case is different: 
a quark can radiate gluons diffractively in the forward direction. 
This happens due to a possibility of interaction between the 
radiated gluon and the target. Such a process, in particular, 
becomes important in diffractive heavy flavor production \cite{KPST07}.
\begin{figure}[h!]
\centerline{\epsfig{file=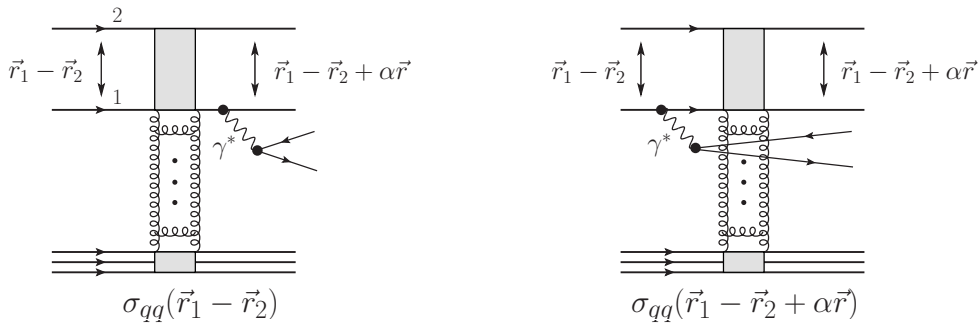,width=13cm}} \caption{Leading
order contribution to the diffractive Drell-Yan in the dipole-target
collision.} \label{fig:gam}
\end{figure}

This is different for the boson radiation off a dipole Fig.~\ref{fig:gam}. 
Such radiation induces a change in transverse separation between
the dipole constituents after the scattering. Since different-size dipoles 
interact with the target with a different strength the diffractive radiation 
amplitude in this case is given by a difference \cite{KPST06}
\begin{eqnarray}
 M^{\rm diff}_{\bar q q} \propto \Psi(\alpha,\vec{r}) 
 \Big(2\mathrm{Im}\, f_{\rm el}(\vec{b},\vec{R}) - 
 2\mathrm{Im}\, f_{\rm el}(\vec{b},\vec{R}+\alpha\vec{r})\Big) \,,
 \label{amp-LO}
\end{eqnarray}
where $\Psi_{q\to G^*q}$ is the light-cone (non-normalised) wave 
function of $q\to G^*q$ fluctuation corresponding to bremsstrahlung 
of virtual gauge bosons $G=\gamma,\,Z,\,W^\pm$ of mass $M$ \cite{Pasechnik:2012ac}, 
$\vec R=\vec r_1-\vec r_2$ is the transverse size of the $q\bar q$ dipole, 
$\alpha$ is the momentum fraction of the gauge boson $G$ 
taken off the parent quark $q$ and $r\sim 1/M$ is the hard scale.

When applied to diffractive $pp$ scattering the diffractive amplitude 
(\ref{amp-LO}), thus, occurs to be sensitive to the large transverse 
separations between the projectile quarks in the incoming proton. 
Normally, transition to the hadron level is achieved by using the
initial proton $\Psi_i$ and remnant $\Psi_f$ wave functions
which encode information about distributions of consituents. 
The completeness relation reads
\begin{eqnarray}\nonumber
&&\sum_{fin}\Psi_{fin}(\vec{r}_1,\vec{r}_2,\vec{r}_3;\{x_q^{1,2,...}\},\{x_g^{1,2,...}\})
\Psi^*_{fin}(\vec{r}\,'_1,\vec{r}\,'_2,\vec{r}\,'_3;\{{x'}_q^{1,2,...}\},\{{x'}_g^{1,2,...}\})\\
&&\phantom{.......}=\,
\delta\bigl(\vec{r}_1-\vec{r}\,'_1\bigr)\delta(\vec{r}_2-\vec{r}\,'_2)
\delta(\vec{r}_3-\vec{r}\,'_3)\prod_{j}\delta(x_{q/g}^j-{x'}_{q/g}^j) \,.
\end{eqnarray}
Here, $\vec r^{\,i}_{q/g}$, $x_{q/g}^i$ are the coordinates and fractions 
of the valence and sea partons, respectively.

Since gluons and sea quarks are mostly accumulated in a close vicinity of valence quarks 
(inside gluonic ``spots''), to a reasonable accuracy the transverse positions of sea quarks 
and gluons can be identified with the coordinates of valence quarks. The valence part of 
the wave function is often taken to be a Gaussian distribution such that
\begin{eqnarray}
\nonumber
|\Psi_{in}|^2&=&\frac{3a^2}{\pi^2}
e^{-a(r_1^2+r_2^2+r_3^2)}\;{\cal R}\big(x_1,\{x_q^{1,2,...}\},\{x_g^{2,3,...}\}\big) \\
&\times&\delta(\vec{r}_1+\vec{r}_2+\vec{r}_3)\delta\Big(1-x_1-\sum_j x_{q/g}^j\Big) \,, \label{psi}
\end{eqnarray}
where all the partons not participating in the hard interaction are summed up; 
$x_1$ is the photon fraction taken from the initial proton; 
$a=\langle r_{ch}^2 \rangle^{-1}$ is the inverse proton 
mean charge radius squared; ${\cal R}$ is a collinear multi-parton distribution in the proton.
Once the latter is integrated over all the partons not participating in the hard interaction,
one gets a conventional collinear PDF $g(x_1,\mu^2)$ for gluons and $q(x_1,\mu^2)$ for 
a given quark flavor $q$. Since the diffractive $pp$ cross section appears as a sum of 
diffractive excitations of the proton constituents, valence/sea quarks and gluons 
are incorporated as
 \beq
 \bigl|\Psi_{in}(\vec r_i,x_i)\bigr|^2 \propto
 {1\over3}\left[\sum\limits_q q(x)+\bar q(x) +
 {81\over16}\,g(x)\right]\ , \label{PDFd}
 \eeq
after intergation over spectator impact parameters and momentum 
fractions with a proper color factor between quark and gluon PDFs.
Note, only sea and valence quarks are excited by the photon
radiation in the diffractive DY process which provide a direct access
to the proton structure function in the soft limit of large $x$ \cite{KRT00}
\[
\sum_q Z_q^2 [q(x)+\bar q(x)]=\frac{1}{x}\,F_2(x)\,.
\]
For diffractive gluon radiation one should
account for both quark and gluon excitations whose amplitudes, however,
are calculated in different ways \cite{KST-par}.

Due to the internal transverse motion of the projectile valence quarks 
inside the incoming proton, which corresponds to finite large transverse 
separations between them, the forward photon radiation does not vanish \cite{KPST06,Pasechnik:2012ac}.
These large distances are controlled by a nonperturbative (hadron) scale $\vec R$, such
that the diffractive amplitude has the Good-Walker structure,
\begin{eqnarray}
M^{\rm diff}_{\bar q q} \propto \sigma(\vec R)-\sigma(\vec R-\alpha\vec r) \propto \vec{r}\cdot \vec{R} \,,
\label{break}
\end{eqnarray}
while the single diffractive-to-inclusive cross sections ratio behaves as
\beqn
\frac{\sigma_{sd}^{\rm DY}}{\sigma_{incl}^{\rm DY}}\propto \frac{\exp(-2R^2/R_0^2(x_2))}{R_0^2(x_2)}
\eeqn
assuming the saturated GBW shape of the dipole cross section (\ref{fel}) where $x_2$
is defined in Eq.~(\ref{x12}). Thus, the soft part of the interaction is not enhanced in Drell-Yan diffraction which
is semi-hard/semi-soft like inclusive DIS. Linear dependence on the hard scale 
$r\sim 1/M\ll R_0(x_2)$ means that even at a hard scale the Abelian radiation
is sensitive to the hadron size due to a dramatic breakdown of diffractive 
factorization \cite{Landshoff-DDY}. It was firstly found in Refs.~\cite{Collins93,Collins97} 
that factorization for diffractive Drell-Yan reaction fails due to the presence of
spectator partons in the Pomeron. In Refs.~\cite{KPST06,Pasechnik:2011nw,Pasechnik:2012ac} it was
demonstrated that factorization in diffractive Abelian radiation is thus 
even more broken due to presence of spectator partons in the colliding
hadrons as reflected in Eq.~(\ref{break}).

\begin{figure}[h]
 \includegraphics[width=8cm]{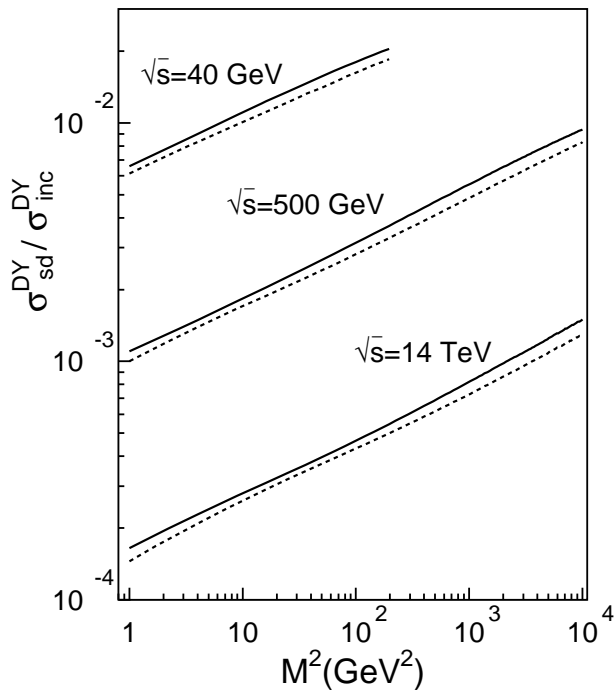}
\caption{The single diffractive-to-inclusive DY cross sections ratio
as a function of the photon virtuality $M^2$ for $x_1=0.5$ (solid lines) and $x_1=0.9$
(dashed lines) and c.m.s. energies $\sqrt{s}=40\GeV,\ 500\GeV$ and $14\TeV$ 
(from top to bottom) \cite{KPST06}.}
 \label{fig:ratio}
 \end{figure}

The effect of diffractive factorisation breaking manifests itself in specific features of observables 
like a significant damping of the cross section at high $\sqrt{s}$ compared to the 
inclusive production case as illustrated in Fig.~\ref{fig:ratio}. This is rather unusual, since a diffractive cross section,
which is proportional to the dipole cross section squared, could be
expected to rise with energy steeper than the total inclusive cross
section, like it occurs in the diffractive DIS process. At the same
time, the ratio of the DDY to DY cross sections was found in
Ref.~\cite{KPST06,Pasechnik:2011nw} to rise with the hard scale, the photon
virtuality $M^2$ also shown in Fig.~\ref{fig:ratio}. This is also in variance with diffraction in DIS, which 
is associated with the soft interactions and where the diffractive 
factorisation holds true \cite{povh}.

Such striking signatures of the diffractive factorisation breaking are due to 
an interplay of soft and hard interactions in the corresponding diffractive 
amplitude. Namely, large and small size projectile fluctuations contribute 
to the diffractive Abelian radiation process on the same footing providing 
the leading twist nature of the process, whereas diffractive DIS dominated 
by soft fluctuations only is of the higher twist \cite{KPST06,Pasechnik:2011nw}.
But this is not the only source of the factorisation breaking -- another important 
source is the absorptive (or unitarity) corrections.

\subsection{Gap survival amplitude}

In the limit of unitarity saturation (the so-called black disk limit) the absorptive 
corrections can entirely terminate the large rapidity gap process. The situation
close to this limit, in fact, happens in high energy (anti)proton-proton collisions
such that unitarity is nearly saturated at small impact parameters \cite{Kopeliovich:2000ef-1,Kopeliovich:2000ef-2}.
The unitarity corrections are typically parameterized by a suppression factor also 
known as the soft survival probability which significantly reduce the diffractive cross section.
In hadronic collisions this probability is controlled by the soft spectator partons 
which are absent in the case of diffractive DIS causing the breakdown 
of diffractive factorisation.

It is well-known that the absorptive corrections affect differently the diagonal and
off-diagonal terms in the hadronic current \cite{KPSdiff,PCAC}, in opposite
directions, leading to an additional source of the QCD factorisation breaking
in processes with off-diagonal contributions only. Namely, the absorptive 
corrections enhance the diagonal terms at larger $\sqrt{s}$, whereas they 
strongly suppress the off-diagonal ones. In the diffractive DY process a
new state, the heavy lepton pair, is produced, hence, the whole
process is of entirely off-diagonal nature, whereas the
diffractive DIS process contains both diagonal and off-diagonal
contributions \cite{KPSdiff}.

The amplitude Eq.~(\ref{amp-LO}) implicitly incorporates the absorptive effects
thus does not requires a soft survival probability multiplier like traditionally imposed 
\cite{Pasechnik:2012ac}. Consider a naive example when a dipole scatters elastically 
off a given potential. Then, the corresponding dipole partial amplitude emerges in 
the following eikonal form
\begin{equation}
\mathrm{Im}\,f_{el}(\vec{b},\vec{r}_1-\vec{r}_2)= 1 - \exp
\Big(i\chi(\vec r_1) - i\chi(\vec r_2)\Big) \,, \qquad 
\chi(b)=-\int\limits_{-\infty}^\infty dz\,V(\vec b,z) \,,
\label{model}
\end{equation}
in terms of the potential $V(\vec b,z)$. This amplitude is close to 
imaginary in the high-energy limit. A diffractive amplitude is then always
proportional to the following difference
\begin{equation}
\mathrm{Im}\,f_{\rm el}(\vec{b}, \vec{r}_1-\vec{r}_2+\alpha\vec r) - \mathrm{Im}\,f_{\rm el}(\vec{b}, \vec{r}_1-\vec{r}_2)\simeq
\exp\Big(i\chi(\vec r_1)-i\chi(\vec r_2)\Big)\, \exp\Big(i\alpha\,\vec r\cdot\vec\nabla\chi(\vec r_1)\Big)\,,
\label{diff}
\end{equation}
where the first exponential factor provides the survival amplitude vanishing in the limit of
the black disc as needed such that the diffractive 
amplitude Eq.~(\ref{amp-LO}) incorporates all absorptive
corrections (gap survival amplitude), provided that the dipole cross section
is adjusted to the data. While normally the survival factor is incorporated into the diffractive observables 
probabilistically, Eq.~(\ref{amp-LO}) treats more naturally quantum-mechanically.

The diffractive gluon radiation is know to be rather weak (the
3-Pomeron coupling is small). This phenomenological observation 
can be explained assuming that gluons in the proton are predominantly 
located inside small ``gluonic spots'' of size $r_0\sim 0.3\fm$ around the
valence quarks (see e.g. Refs.~\cite{KST-par,shuryak,Shuryak:2003rb,spots}). 
The smallness of gluonic dipole is an important nonperturbative phenomenon
which may be connected e.g. to the small size of gluonic fluctuations
in the instanton liquid model \cite{Shuryak:2003rb}.
Therefore, a distance between a valence quark and a gluon in a vicinity of
another quark can be safely approximated by the
quark-quark separation.

Besides the soft gluons in the proton light-cone wave function, virtual
gauge boson production triggers intensive gluon radiation such that 
there are many more spectator gluons in a vicinity of the quark which radiates 
the gauge boson. The separations of such gluons from the parent quark are 
controlled by the QCD DGLAP dynamics. In practice, one may replace such a set 
of gluons by dipoles \cite{mueller} whose transverse sizes $r_d$ vary between 
$1/M_G$ and $r_0$ scales \cite{nz94}. Then the mean dipole
size is regulated by a relation 
 \beq 
 \la r_d^2\ra =\frac{r_0^2}{\ln(r_0^2M_G^2)} \,, \label{mean-dip} 
 \eeq 
leading to $\la r_d^2\ra\approx 0.01\fm^2$, which means that it is
rather small and the corresponding dipole cross section 
$\sigma \simeq C(x)\,\la r_d\ra^2$, where $C(x)=\sigma_0/R_0^2(x)$ 
rises with energy, is suppressed. For $x=M_G^2/s$ and naive GBW parameterisation 
\cite{GBWdip} we get $\sigma\approx 0.9\mb$ at the Tevatron energy.
Each such small dipole brings up an extra suppression factor to the large rapidity gap 
survival amplitude given by 
 \beq 
 S_d(s)=1-\Im f_d(b,r_d) \,. \label{S_d} 
 \eeq 
Here, the second term is small and thus is simplified to (for more details, see Ref.~\cite{dis}), 
 \beq
 \Im f_d(b,r_d)\approx \frac{\sigma_d}{4\pi B_d}\,e^{-b^2/2B_d} \,, \label{f_d}
 \eeq
where $B_d$ is the standard dipole-nucleon elastic slope $B_d\approx 6\GeV^{-2}$ measured 
earlier in diffractive $\rho$ electro-production at HERA \cite{rho}.
At the mean impact parameter given by $\la b^2\ra=2B_d$ and for the Tevatron energy
$\sqrt{s}=2\TeV$ we arrive at negligibly small value for the absorptive 
correction (\ref{f_d}): $\Im f_d(0,r_d)\approx 0.01$. 

On the other hard, the overall 
number of such dipoles increases with hardness of the process, which can amplify 
the magnitude of the absorptive effect. Generalising the gap survival amplitude to $n_d$
projectile dipoles, we obtain
 \beq S^{(n_d)}_d=\bigl[1-\Im f_d(b,r_d)\bigr]^{n_d} \,. 
 \label{S_n} 
 \eeq
The DGLAP evolution formulated in impact parameter representation \cite{nz94}
enables to estimate the mean number of such dipoles can be estimated in the
double-leading-log approximation
 \beq 
 \la n_d\ra = \sqrt{\frac{12}{\beta_0}
 \ln\left(\frac{1}{\alpha_s(M_G^2)}\right) \ln\left((1-x_F){s\over s_0}\right) } \,. 
 \label{mean-n} 
 \eeq 
Here, the typical Bjorken $x$ values of the radiated gluons is restricted by the diffractive mass as 
$x>s_0/M_X^2=s_0/(1-x_F)s$. In typical kinematics at the Tevatron collider, the mean number of such
dipoles is roughly $\la n_d\ra\lesssim6$. The amplitude of survival of a large rapidity 
gap is controlled by the largest dipoles in the projectile hadron only, such that the first 
exponential factor in Eq.~(\ref{diff}) provides a sufficiently good approximation to the 
gap survival amplitude. The absorptive corrections (\ref{S_n}) to the gap survival 
amplitude are proven to be rather weak and do not exceed $5\%$ (or $10\%$ in the survival
probability factor) which is small compared to an overall theoretical uncertainty.
For the pioneering work on hard rescattering corrections to the gap survival factor 
see Ref.~\cite{Bartels:2006ea}.

The popular quasi-eikonal model for the so-called ``enhanced'' probability
$\hat{S}_{enh}$ (see e.g. Refs.~\cite{enh-1,enh-2}), frequently used 
to describe the factorisation breaking in diffractive processes, is not well 
justified in higher orders, whereas the color dipole approach considered 
here, correctly includes all diffraction excitations to all orders
\cite{KPSdiff}. Such effects are included into the phenomenological 
parameterizations for the partial elastic dipole amplitude fitted to data. 
This allows to predict the diffractive gauge bosons production cross sections 
in terms of a single parameterization for the universal dipole cross section 
(or, equivalently, the elastic dipole amplitude) known independently from 
the soft hadron scattering data. 

For more details on derivations of diffractive gauge boson production amplitudes 
and cross sections see Refs.~\cite{Pasechnik:2011nw,Pasechnik:2012ac}. 
Now we turn to a discussion of numerical results for the most important observables.

\section{Single diffractive gauge bosons production}
\label{numerics}

\begin{figure*}[!h]
\begin{minipage}{0.49\textwidth}
 \centerline{\includegraphics[width=1.0\textwidth]{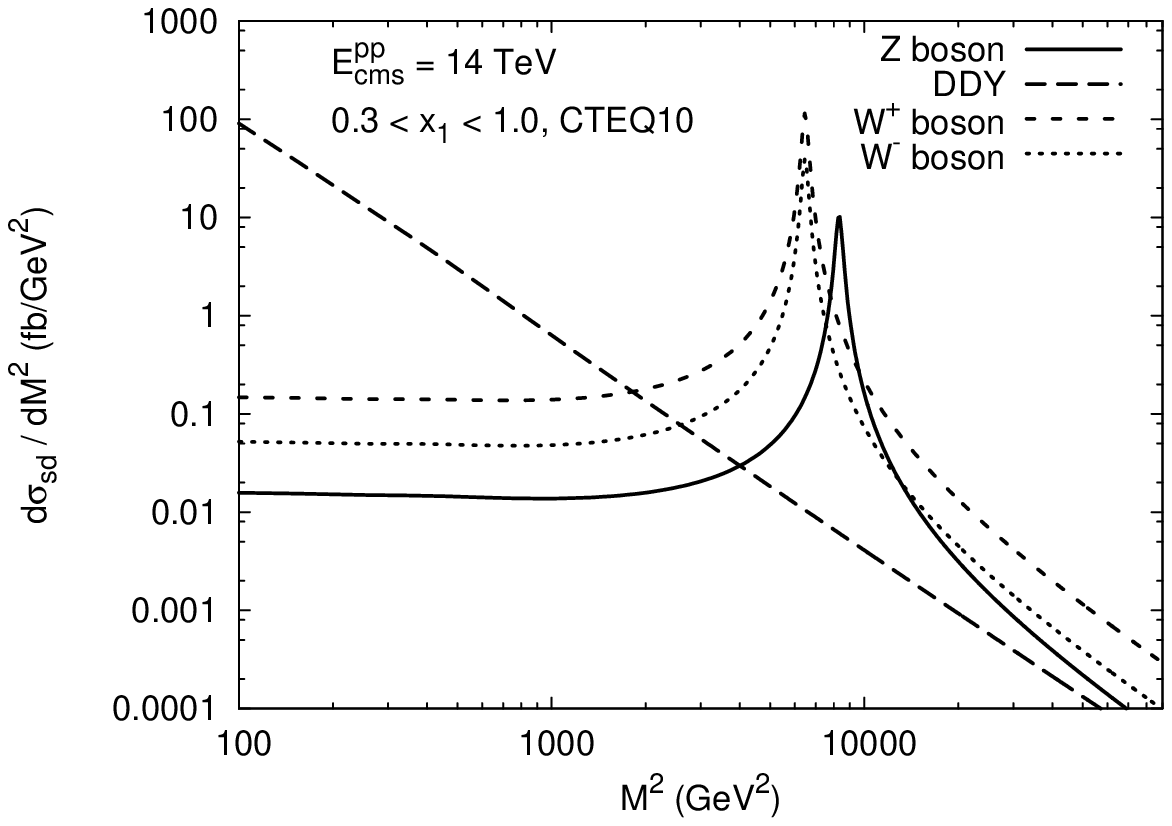}}
\end{minipage}
\hspace{0.5cm}
\begin{minipage}{0.46\textwidth}
 \centerline{\includegraphics[width=1.0\textwidth]{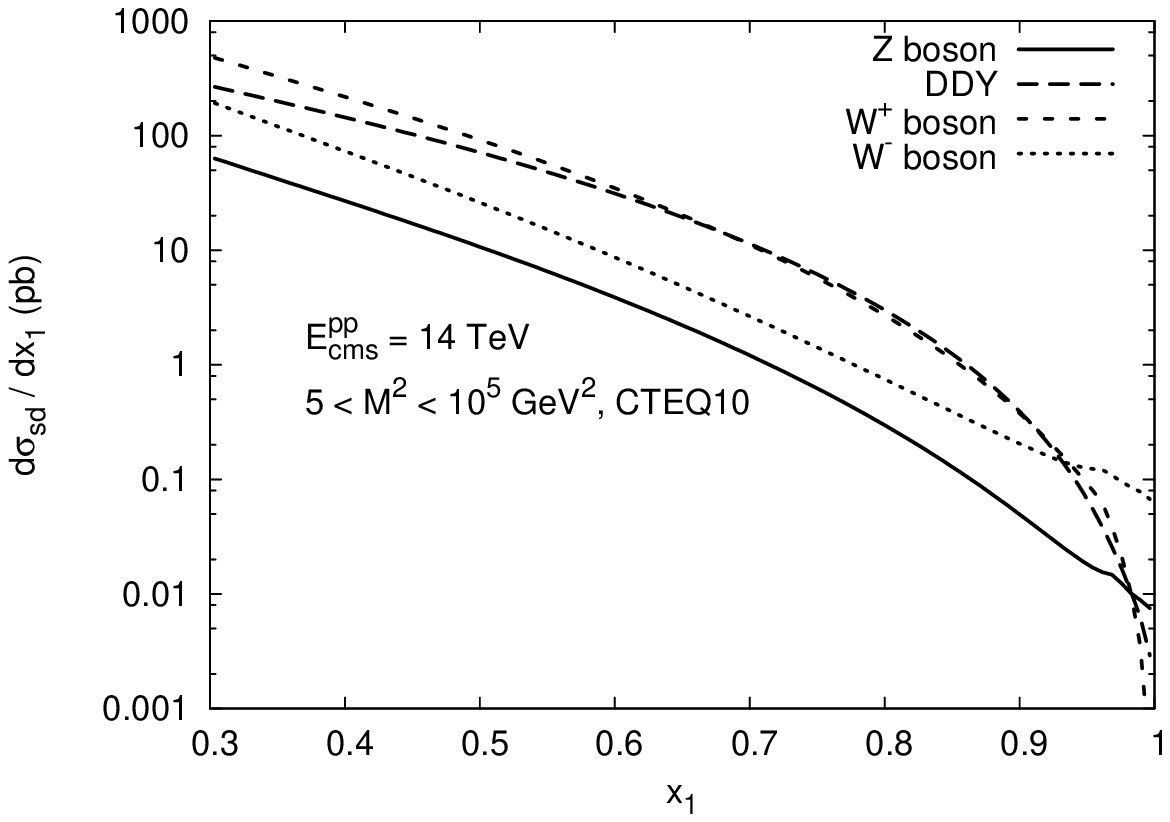}}
\end{minipage}
   \caption{
\small 
The cross section for diffractive boson production as a function
of $M^2$ (left) and fraction $x_1$ (right) at the energy of LHC.}
 \label{fig:CS-LHC}
\end{figure*}

In Ref.~\cite{Pasechnik:2012ac} the dipole framework has been used in
analysis of diffractive gauge bosons production, and here we briefly
overview these results. The corresponding observables for $Z^0,\,\gamma^*,\,W^{\pm}$ 
production ($\sqrt{s}=14$ TeV) such as $d\sigma_{sd}/dM^2$ and $d\sigma_{sd}/dx_1$ are shown in Fig.~\ref{fig:CS-LHC}
at left and right panels, respectively. The $M^2$ distributions correspond to
the forward rapidities, i.e. $0.3<x_1<1$ and the interval $5<M^2<10^5$ GeV$^2$ is concerned.

In the region corresponding to the resonant $Z^0,\,W^{\pm}$ bosons production,
the $M^2$ distributions exceed the diffractive $\gamma^*$
component. The latter is relevant for low masses only. 
As for the $x_1$-distributions of $W^+$ and
$\gamma^*$ components, these are relatively close to each other, 
whereas the $W^-,\,Z$-boson components are smaller. 
A precision measurement of diffractive $W^\pm$ distributions and their differences 
may enable further constraining of the quark PDFs at 
large quark momentum fractions $x_1/\alpha$.
\begin{figure}[!h]
\begin{minipage}{0.49\textwidth}
 \centerline{\includegraphics[width=1.0\textwidth]{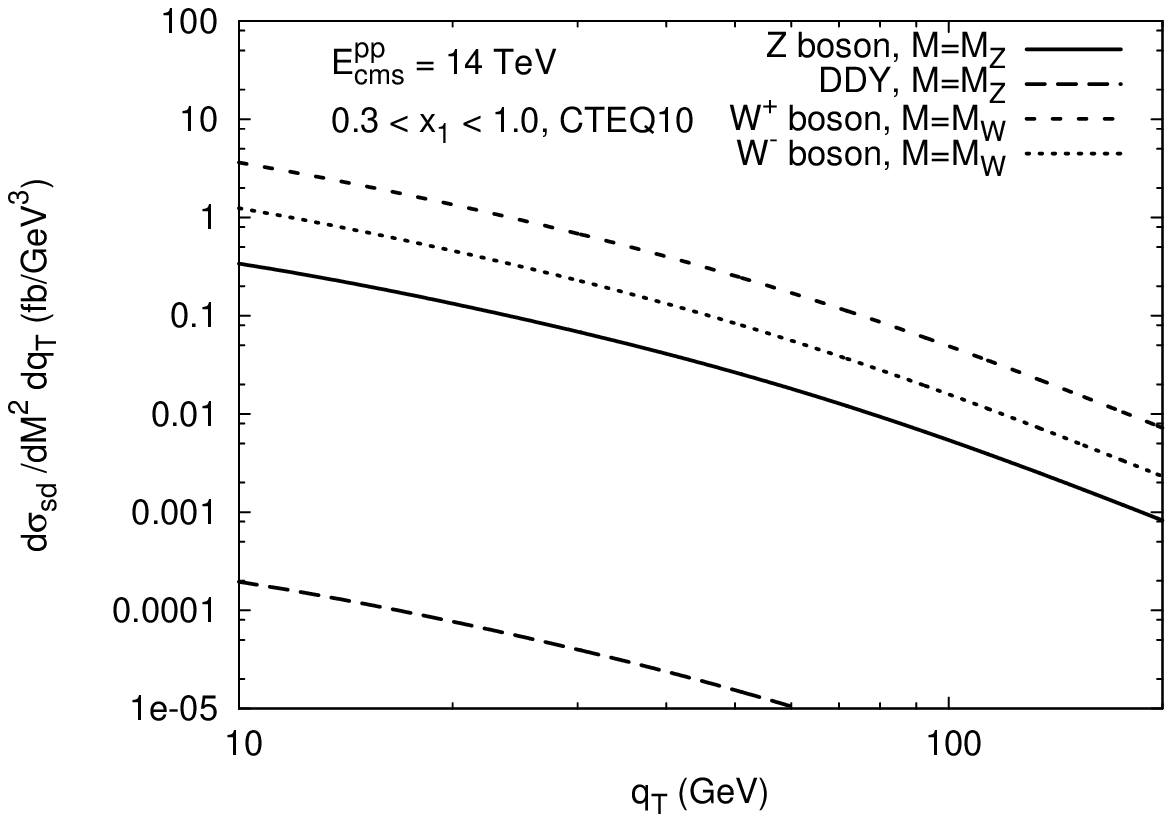}}
\end{minipage}
\hspace{0.5cm}
\begin{minipage}{0.46\textwidth}
 \centerline{\includegraphics[width=1.0\textwidth]{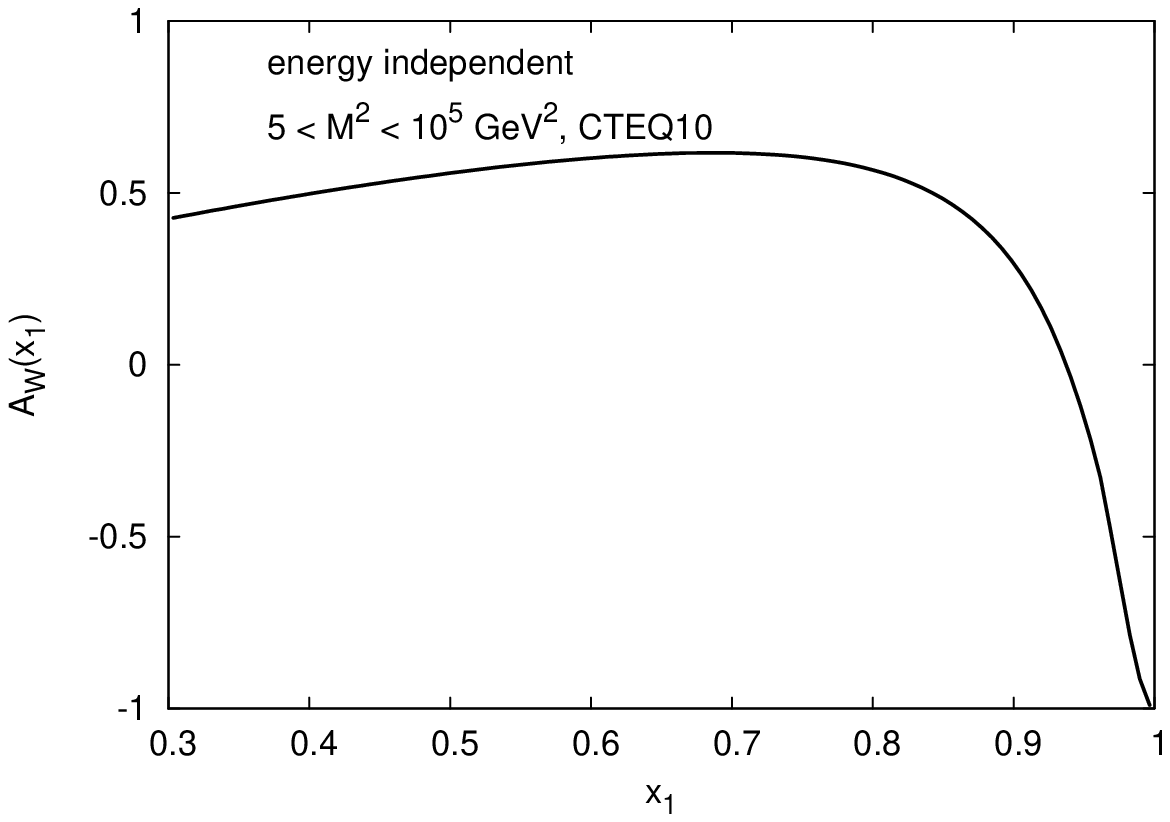}}
\end{minipage}
   \caption{
\small The lepton pair transverse momentum $q_{\perp}$ distribution the 
diffractive cross section at the LHC $\sqrt{s}=14$ TeV (left panel) and
the charge asymmetry in the SD $W^\pm$ cross sections at fixed 
$M^2=M_W^2$ (right panel).}
 \label{fig:dqt-A}
\end{figure}

Another phenomenologically interesting observable is the lepton-pair 
$q_\perp$ differential distribution at the LHC shown in Fig.~\ref{fig:dqt-A} (left). 
The $W^{\pm}$ charge asymmetry
is particularly dependent on the $u,\,d$ PDFs difference at large $x$. It is given by
\begin{eqnarray}
 A_W \equiv \frac{d\sigma_{sd}^{W^+}/dx_1 -
 d\sigma_{sd}^{W^-}/dx_1}{d\sigma_{sd}^{W^+}/dx_1
 +d\sigma_{sd}^{W^-}/dx_1} \,.
\end{eqnarray}
and is shown in Fig.~\ref{fig:dqt-A} (right panel). The quantity does not depend on 
both the invariant mass/energy. 
\begin{figure*} [!h]
\begin{minipage}{0.49\textwidth}
 \centerline{\includegraphics[width=1.0\textwidth]{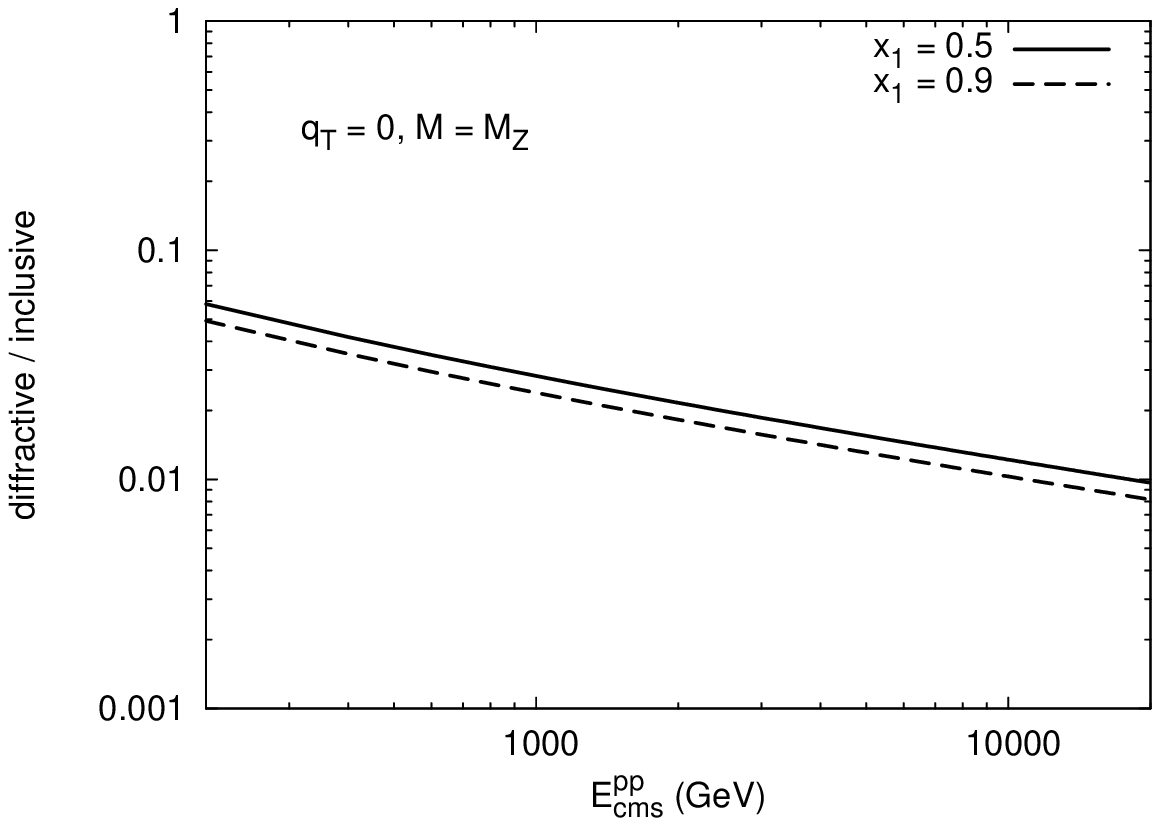}}
\end{minipage}
\hspace{0.2cm}
\begin{minipage}{0.48\textwidth}
 \centerline{\includegraphics[width=1.0\textwidth]{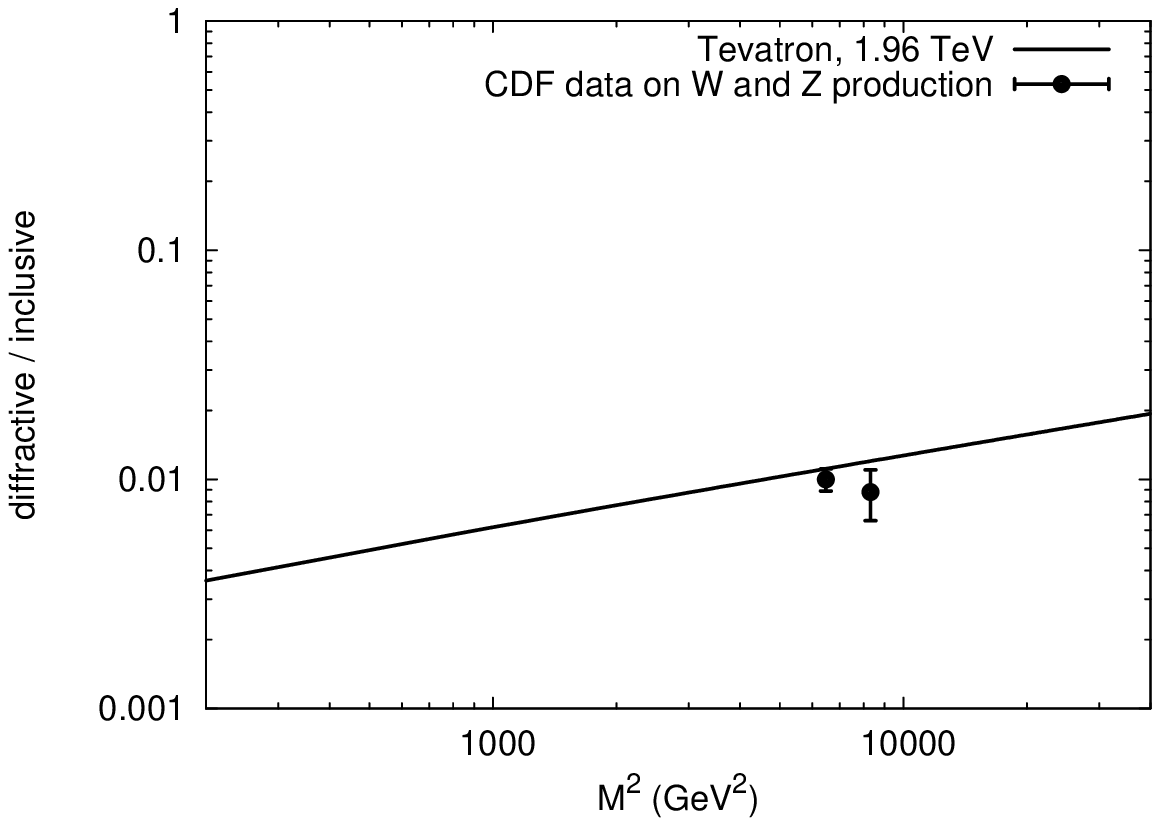}}
\end{minipage}
   \caption{
\small The SD-to-incl. ratio w.r.t. the collision energy
 $\sqrt{s}$ (left) as well as $M^2$ at Tevatron (right).} \label{fig:ratio1}
\end{figure*}

Similarly to diffractive DY discussed above, an important feature of 
the SD-to-inclusive ratio as a function of $M^2,\,x_1$
\begin{eqnarray}\label{Rrat}
R=\frac{d\sigma_{sd}/dx_1dM^2} {d\sigma_{incl}/dx_1dM^2}\,,
\end{eqnarray}
which exhibits an non-typical energy as well as hard scale dependence (see Fig.~\ref{fig:ratio1})
compared to the conventional diffractive QCD factorisation based predictions 
\cite{Szczurek, Beatriz}. In analogy to DDY case, this ratio behaves w.r.t. the energy 
and the hard scale in opposite way to what is expected from diffractive factorisation.
The ratio does not depend on the properties of the radiated gauge boson and 
PDFs while it is sensitive to the partial dipole amplitude structure only efficiently probing 
the QCD mechanism of diffraction. Thus, the diffractive gauge boson 
observables in the di-lepton channel which enhanced compared to DDY 
around the $Z^0$ and $W^\pm$ resonances provides crucial details on the 
soft/hard fluctuations and their interplay in QCD.

\section{Diffractive non-Abelian radiation}

As we have seen in the discussion above, diffractive DY is one of the
most important examples of leading-twist processes, where simultaneously 
large and small size projectile fluctuations are at work. It turns out that 
the participation of soft spectator partons in the interaction with 
the gluonic ladder is crucial and results in a leading twist effect.
What are other examples of the leading twist behavior in diffraction?

\subsection{Leading-twist diffractive heavy flavor production}

One might naively think that the Abelian (or DIS) mechanism of heavy flavor production 
$\gamma^*\to Q\bar Q$ is of the leading twist as well since it behaves
as $\sim 1/Q^2$. However, in the limit $m_Q^2\gg Q^2$ the corresponding
cross section $\sigma_{sd}\propto 1/m_Q^4$ i.e. behaves as a higher
twist process. One has to radiate at least one gluon off the $Q\bar Q$ pair
for this process to become the leading twist one, e.g. 
$\sigma_{sd}(\gamma^*\to Q\bar Qg)\propto 1/m_Q^2$, since the mean
transverse separation between $G$ and small $Q\bar Q$ dipole is typically large
although formally such a process is of the higher perturbative QCD order in $\alpha_s$.
\begin{figure*} [!h]
 \centerline{\includegraphics[width=0.7\textwidth]{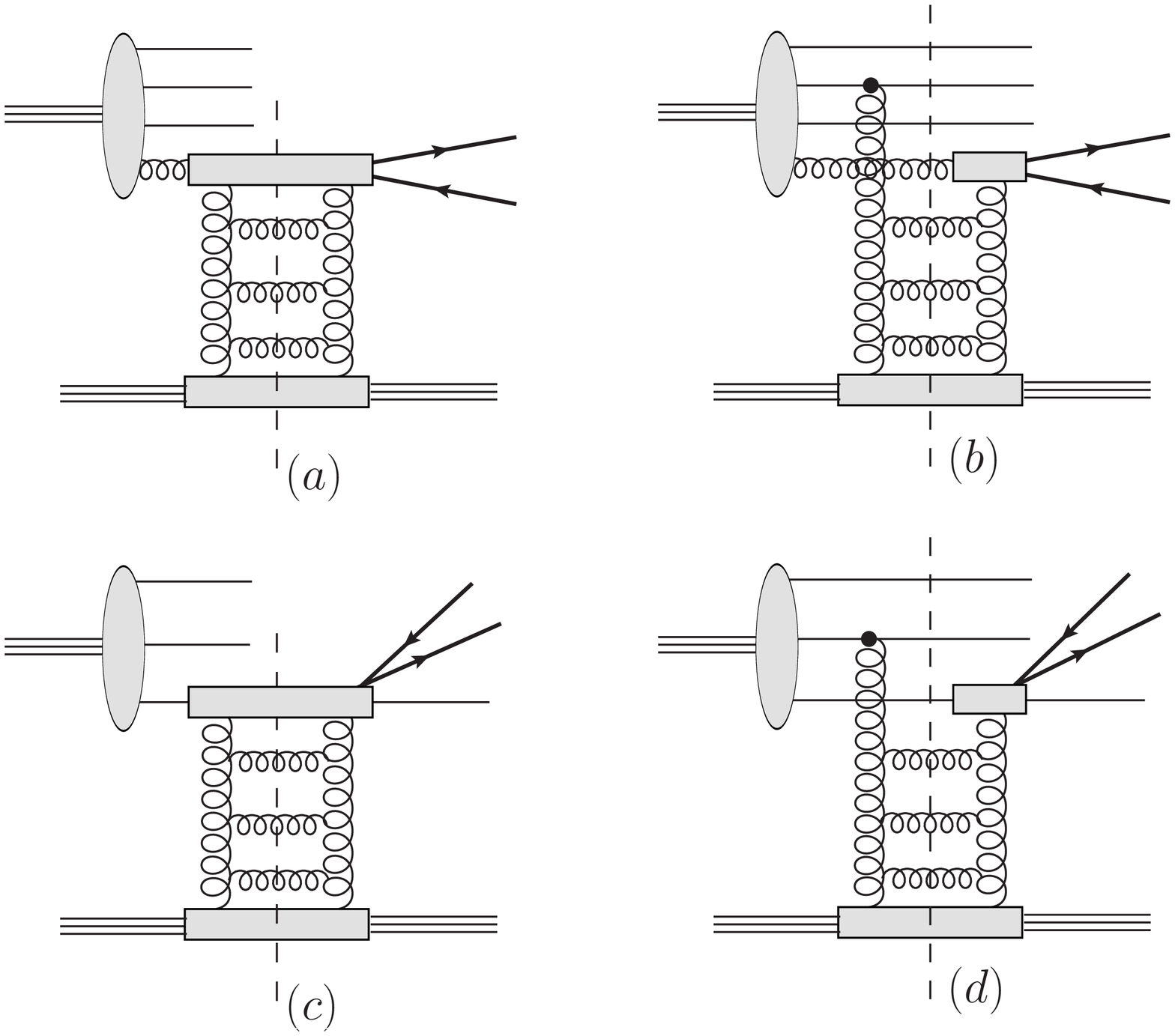}}
   \caption{
\small Leading order contributions to single diffractive heavy flavor
production in gluon-proton (a,b) and quark-proton 
(c,d) scattering subprocesses in $pp$ collisions. Diagrams (b,d)
emerge due to the presence of soft spectator partons in the proton
wave function (the screening gluon couples to every spectator
parton besides the active one). Grey effective vertices account for
all possible couplings of the incident partons.}
 \label{fig:HF}
\end{figure*}

Consider now the non-Abelian mechanism for diffractive hadroproduction of heavy 
quarks via $g^*\to Q\bar Q$ hard subprocess. Production of heavy quarks at large
$x_F\to 1$ is a longstanding controversial issue even in inclusive processes. On one hand,
QCD factorisation approach predicts vanishingly small yields of heavy flavor due to
steeply falling gluon density as $\sim (1-x_F)^5$ at large $x_F$. On the other hand,
the end-point behavior is controlled by the universal Regge asymptotics 
$d\sigma/dx_F(x_F\to 1)\propto (1-x_F)^{1-2\alpha_R(t)}$ in terms of the Regge
trajectory of the $t$-channel exchange $\alpha_R(t)$. Apparently, the Regge and
QCd factorisation approaches contradict each other. The same problem emerges
in the DY process at large $x_F$ as is seen in data \cite{Wijesooriya:2005ir} which means that in 
the considering kinematics the conventional QCD factorisation does not apply \cite{Kopeliovich:2005ym}.
At the same time, the observation of an excess of diffractive production of heavy quarks
at large $x_F\to1$ compared to conventional expectation may provide a good evidence 
for intrinsic heavy flavors if the latter is reliably known. Calculations assuming that diffractive 
factorisation holds for hard diffraction \cite{ISh,Donnachie:1987gu} may not be used for 
quantifying the effect from intrinsic heavy flavor. Instead, the dipole framework has been 
employed to this process for the first time in Ref.~\cite{KPST07}. Here we briefly overview 
the basic theory aspects concerning primarily heavy quarks produced in the projectile fragmentation
region (for inclusive $Q\bar Q$ production at mid rapidites in the dipole framework, 
see Ref.~\cite{Kopeliovich:2002yv}).

Typical contributions to the single diffractive $Q\bar Q$ production rate are summarized in Fig.~\ref{fig:HF}. 
Diagrams (a) and (b) correspond to the leading order gluon splitting into $Q\bar Q$ contributions 
in the color field of the target proton (diffractive gluon excitation). The latter gluon as 
a component of the projectile proton wave function can be treated as real 
(via collinear gluon PDF) or virtual (via unintegrated 
gluon PDF). Due to hard scale $m_Q$ the diagram (a) with Pomeron coupling to a
small-size $gQ\bar Q$ system is of the higher twist due to color transparency
and is therefore suppressed. Diagram (b) involves two scales -- the soft hadronic one
$\sim \Lambda_{\rm QCD}$ associated with large transverse separations between
a gluon and constituent valence quarks, and the hard one $\sim m_Q$ associated with
small $Q\bar Q$ dipole. An interplay between these two scales similar to that in DDY
emerges as the leading twist effect; thus, diagram (b) is important. Possible higher order
terms with an extra gluon radiation contributing to the leading twist diffractive heavy flavor 
production were disscussed in detail in Ref.~\cite{KPST07}.

\begin{figure*} [!h]
 \centerline{\includegraphics[width=0.7\textwidth]{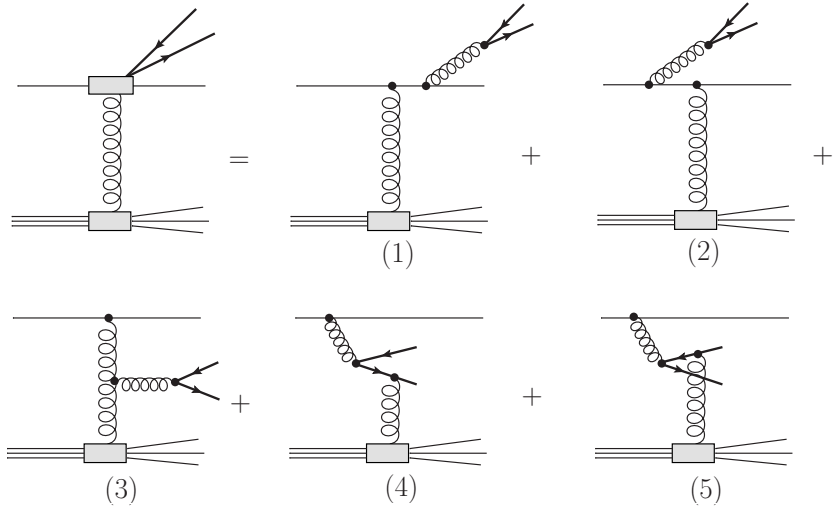}}
   \caption{
\small Five different topologies contributing to inclusive $Q\bar Q$ production
in quark-proton scattering. These can be split into two gauge-invariant subsets
of amplitudes as described in the text.}
 \label{fig:HF-2}
\end{figure*}

Diagrams (c) and (d) account for $Q\bar Q$ production via diffractive quark excitation.
Just as in leading twist diffraction in DIS $\gamma^*\to Q\bar Qg$, these processes 
are associated with two characteristic transverse separations, a small one, $\sim 1/m_Q$, between 
the $\bar Q$ and $Q$, and a large one, either $\sim 1/m_q$ between $q$ and $Q \bar Q$ (diagram (c)) 
or $1/\Lambda_{\rm QCD}$ between another constituent valence quark and $Q \bar Q$ (diagram (d)).
While all the terms contributing to (d) are of the leading twist (see Ref.~\cite{KPST07}), only a special
subset of diagrams (c) are of the leading twist. Indeed, the hard subprocess $q+g\to (Q\bar Q)+q$ 
is characretized by five distinct topologies illustrated in Fig.~\ref{fig:HF-2}, and similar graphs are for
gluon-proton scattering with subprocess $g+g\to (Q\bar Q)+g$.

These graphs can be grouped into two amplitudes attributed to bremsstrahlung (BR) and production (PR) 
mechanisms, which do, or do not involve the projectile light quarks or gluons, respectively 
(for more details, see Fig.~2 and Appendix A in Ref.~\cite{KPST07}). The BR mechanism 
includes the same graphs as radiation of a gluon (see Refs.~\cite{gb,kst1}), i.e. interaction 
with the source parton before and after radiation, and interaction with the radiated gluon. 
The PR mechanism, responsible for the transition $g\to\bar QQ$, includes the interactions with 
the gluon and the produced $\bar QQ$ (also known as gluon-gluon fusion $gg\to Q\bar Q$ mechanism). 
The total amplitude is
 \beq
M_{q,g}=M_{q,g}^{\rm BR} + M_{q,g}^{\rm PR}\,,
 \eeq
where subscripts $q,\,g$ denote contributions with hard gluon
radiation by the projectile valence or sea quarks and gluons,
respectively.  Such grouping is performed separately for transversely
and longitudinally polarised gluons as described in Ref.~\cite{KPST07}). 
One of the reasons for this grouping is that each of these 
two combinations is gauge invariant and can be expressed in terms of three-body 
dipole cross sections, $\sigma_{g\bar qq}$ and $\sigma_{g\bar QQ}$ respectively.

Another physical reason for such a separation is different scale dependence 
of the BR and PR components. Introducing the transverse separations $\vec r,\,\vec r_1$ 
and $\vec r_2$ within the $\bar Q Q$, $q\bar Q$ and $qQ$ pairs, respectively, the
three body dipole cross sections can be expressed via two scales: the distance between the
final light quark (or gluon) and the center of gravity of the $Q\bar Q$ pair, 
$\vec \rho=\vec r-\beta \vec r_1- (1-\beta)\vec r_2$ ($\beta$ is the heavy quark momentum 
fraction taken from the parent gluon which takes fraction $\alpha$ of the parent parton), 
and the $Q\bar Q$ transverse separation, $\vec s=\vec r_1 - \vec r_2$. The corresponding 
distribution amplitudes of $Q\bar Q$ production in diffractive quark/gluon scattering off proton 
\beqn
A_{\rm BR}\propto \Phi_{\rm BR}(\vec \rho,\vec s)\Sigma_1(\vec \rho,\vec s)\,, \qquad
A_{\rm PR}\propto \Phi_{\rm PR}(\vec \rho,\vec s)\Sigma_2(\vec \rho,\vec s)\,,
\eeqn
are given in terms of the effective dipole cross sections for a colorless $g\bar qq$ 
and $g\bar QQ$ systems, and rather complicated wave functions $\Phi$ of subsequent 
gluon radiation and then its splitting into $\bar QQ$ pair in both cases. In the case of bremsstrahlung, 
both mean separations are controlled by the hard scale such that
\[
A_{\rm BR}\sim \langle \rho^2 \rangle\sim \langle s^2 \rangle\sim \frac{1}{m_Q^2} \,,
\]
thus, the corresponding contribution is a higher twist effect and thus suppressed (note, in the case 
of forward Abelian radiation this contribution is equal to zero). On the contrary, in the production
mechanism only the $\bar QQ$ separation is small, $\langle s^2 \rangle\sim 1/m_Q^2$, the second
scale appears to be soft, $\langle \rho^2 \rangle\sim 1/m_q^2$, leading to the leading twist behavior
\[
A_{\rm PR}\sim \vec s \cdot \vec \rho
\]
in analogy to diffractive DY process. This is a rather nontrivial fact, since in the case of 
the DY reaction such a property is due to the Abelian nature of the radiated particle while
here we consider a non-Abelian radiation. The bremsstrahlung-production interference terms
are of the higher twist and thus are safely omitted.
\begin{figure*} [!h]
\begin{minipage}{0.45\textwidth}
 \centerline{\includegraphics[width=1.0\textwidth]{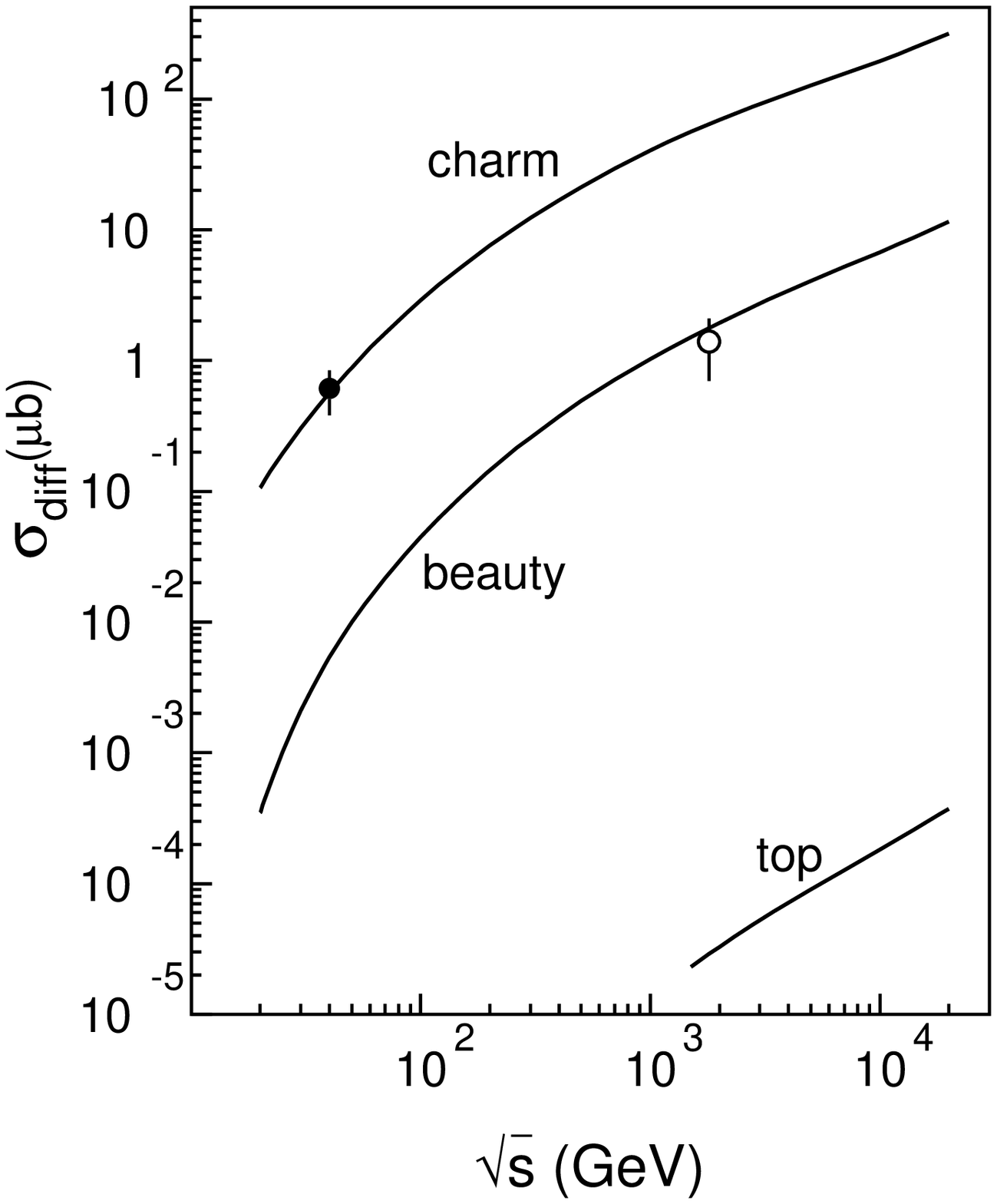}}
\end{minipage}
\hspace{0.2cm}
\begin{minipage}{0.45\textwidth}
 \centerline{\includegraphics[width=1.0\textwidth]{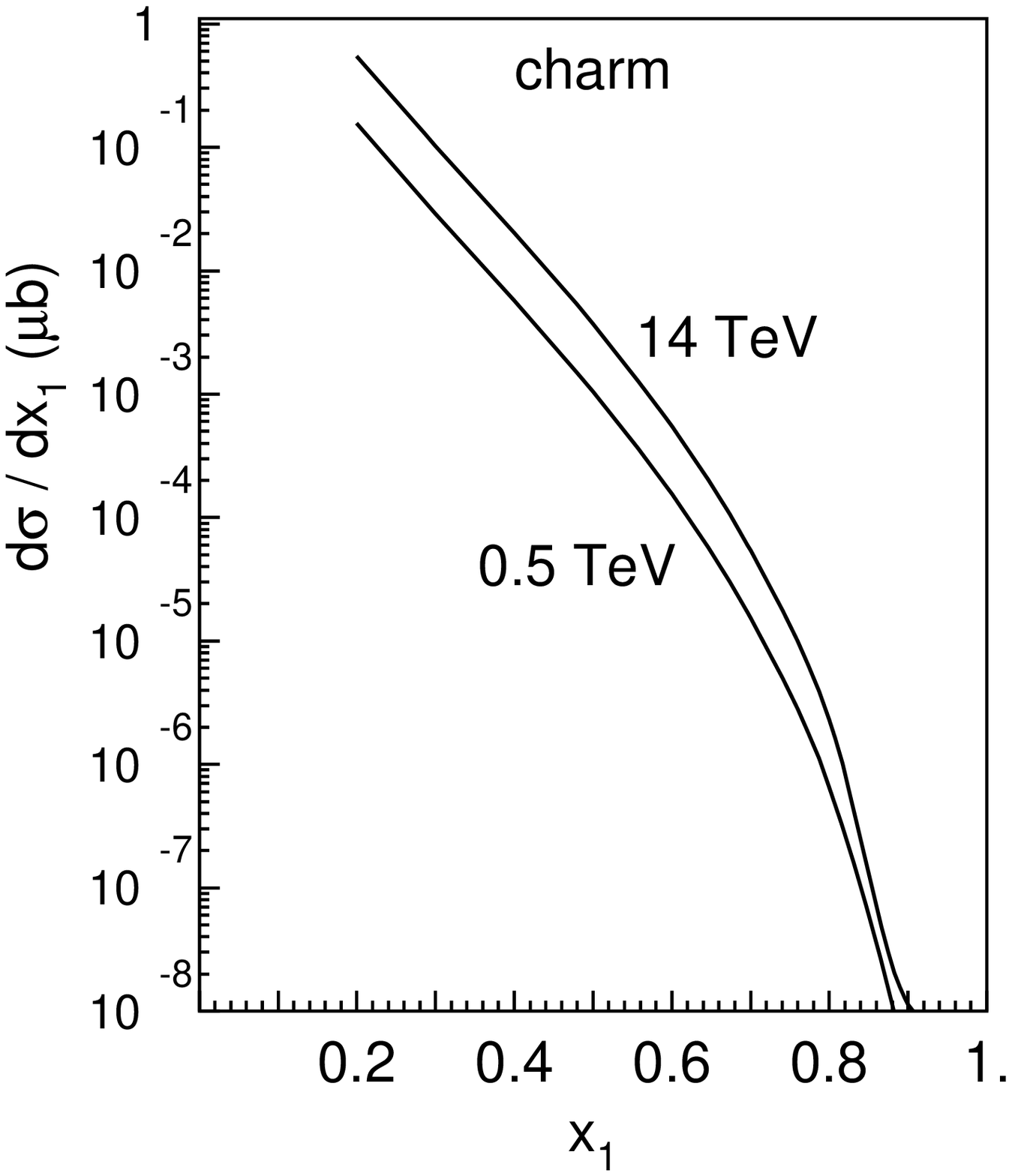}}
\end{minipage}
   \caption{
\small The total cross section of diffractive $c\bar c$, $b\bar b$ and $t\bar t$ pairs production
 as function of energy in comparison with experimental data from E690 \cite{Wang:2000dq} and 
 CDF \cite{Affolder:1999hm} experiments (left panel) and the differential cross section 
 as function of fraction $x_1$ of the initial proton momentum carried by the charm 
 quark (right panel) \cite{KPST07}.}
 \label{fig:HF-3}
\end{figure*}

The situation with scale dependence in the case of $\bar QQ$ production in diffractive $pp$ scattering 
is somewhat similar to diffractive quark-proton scattering discussed above but technically more involved
due to extra terms (b) and (d) in Fig.~\ref{fig:HF} and color averaging over the projectile proton wave 
function. Although bremsstrahlung terms from diagrams (b), (d) are formally of the leading twist due
to interactions with distant spectator partons, numerically they are always tiny due to denominator 
suppression by a large $\bar QQ$ mass. Thus, the leading twist production terms from (b), (c), and 
(d) sets are relevant whereas the set (a) does not contain production terms and is a higher twist effect.
Thus, like in diffractive Drell-Yan in the considering process the leading twist effect, at least, partly
emerges due to intrinsic transverse motion of constituent quarks in the incoming proton. However,
due to a non-Abelian nature of this process extra leading-twist terms production from the ``production'' 
mechanism, which are independent
of the structure of the hadronic wave function, become important. Diffractive production 
cross sections of charm, beauty and top quark pairs, $p+p\to Q\bar QX+p$, as functions 
of c.m.s. $pp$ energy are shown in Fig.~\ref{fig:HF-3}. The experimental data points 
available from E690 \cite{Wang:2000dq} and CDF \cite{Affolder:1999hm} experiments 
have been compared with theoretical predictions evaluated with corresponding phase 
space constraints (for more details, see Ref.~\cite{KPST07}).

\subsection{Single diffractive Higgsstrahlung}

Typically large Standard Model (SM) backgrounds and theoretical uncertainties due to higher order effects 
strongly limit the potential of inclusive Higgs boson production for spotting likely small but yet possible 
New Physics effects. Some of the SM extensions predict certain distortions in Higgs boson Yukawa couplings 
such that the precision multi-channel measurements of the Higgs-heavy quarks couplings becomes a crutial
test of the SM structure. As a very promising but challenging channel, the exclusive and diffractive Higgs 
production processes (involving rapidity gaps) offer new possibilities to constrain the backgrounds, and 
open up more opportunities for New Physics searches (see e.g. Refs.~\cite{Durham-1,Durham-2,Durham-3,
Durham-4,Durham-5,Heinemeyer:2007tu,Heinemeyer:2010gs,Tasevsky:2013iea}). 

The QCD-initiated gluon-gluon fusion $gg\to H$ mechanism via a heavy quark loop is one of the dominant 
and most studied Higgs bosons production modes in inclusive $pp$ scattering which has led to its discovery 
at the LHC (for more information on Higgs physics highlights, see e.g. Refs.~\cite{ATLAS,CMS,Carena:2002es,Handbook-1,Handbook-2,Handbook-3} 
and references therein). The same mechanism is expected to provide an important Higgs production mode 
in single diffractive $pp$ scattering as well as in central exclusive Higgs boson production \cite{Durham-1,Durham-2,Durham-5}. 
The forward inclusive and diffractive Higgsstrahlung off intrinsic heavy flavor at $x_F\to 1$ has previously 
been studied in Refs.~\cite{Brodsky:2007yz,Brodsky:2006wb}, respectively. 

Very recently, a new single diffractive production mode of the Higgs boson in association with a heavy quark 
pair $\bar QQ$, namely $pp\to X+Q\bar QH+p$, at large $x_F$ where conventional factorisation-based approaches 
are expected to fail has been studied in Ref.~\cite{Pasechnik:2014lga}. The latter process is an important background
for diffractive Higgs boson hadroproduction off intrinsic heavy flavor. Here, we provide a short overview 
of this process which is analogical to forward diffractive $\bar QQ$ production discussed above.
\begin{figure*}[!h]
 \centerline{\includegraphics[width=0.7\textwidth]{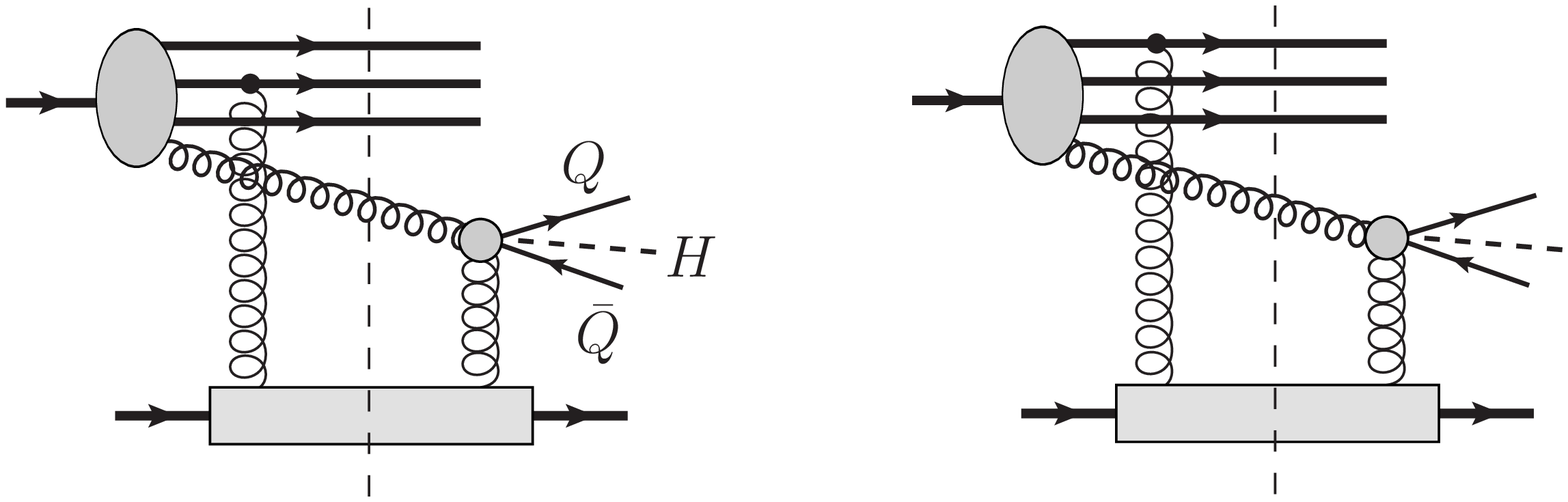}}
   \caption{
\small Dominant gluon-initiated contributions to the single 
diffractive $\bar Q Q + H$ production \cite{Pasechnik:2014lga}.}
 \label{fig:QQh-SD}
\end{figure*}

For a reasonably accurate estimate one retains only the dominant gluon-initiated 
leading twist terms illustrated in Fig.~\ref{fig:QQh-SD} where the ``active'' gluon is coupled to 
the hard $Q\bar Q+H$ system, while the soft ``screening'' gluon couples 
to a spectator parton at a large impact distance. The latter are illustrated by 
tree-level diagrams with Higgs boson radiation off a heavy quark or Higgsstrahlung. 
In practice, however, one does not calculate the Feynman graphs explicitly in Fig.~\ref{fig:QQh-SD}. 
Instead one should adopt the generalized optical theorem within the Good-Walker approach 
to diffraction \cite{GW} such that a diffractive scattering amplitude turns out to be proportional 
to a difference between elastic scatterings of different Fock states \cite{Pasechnik:2014lga}.
The contributions where both ``active'' and ``screening'' gluons couple to partons at small 
relative distances are the higher twist ones and thus are strongly suppressed by extra powers of the hard
scale (see e.g. Refs.~\cite{KPST07}). This becomes obvious 
in the colour dipole framework due to colour transparency \cite{zkl} making the medium 
more transparent for smaller dipoles. 
\begin{figure*}[!ht]
\begin{minipage}{0.495\textwidth}
 \centerline{\includegraphics[width=1.0\textwidth]{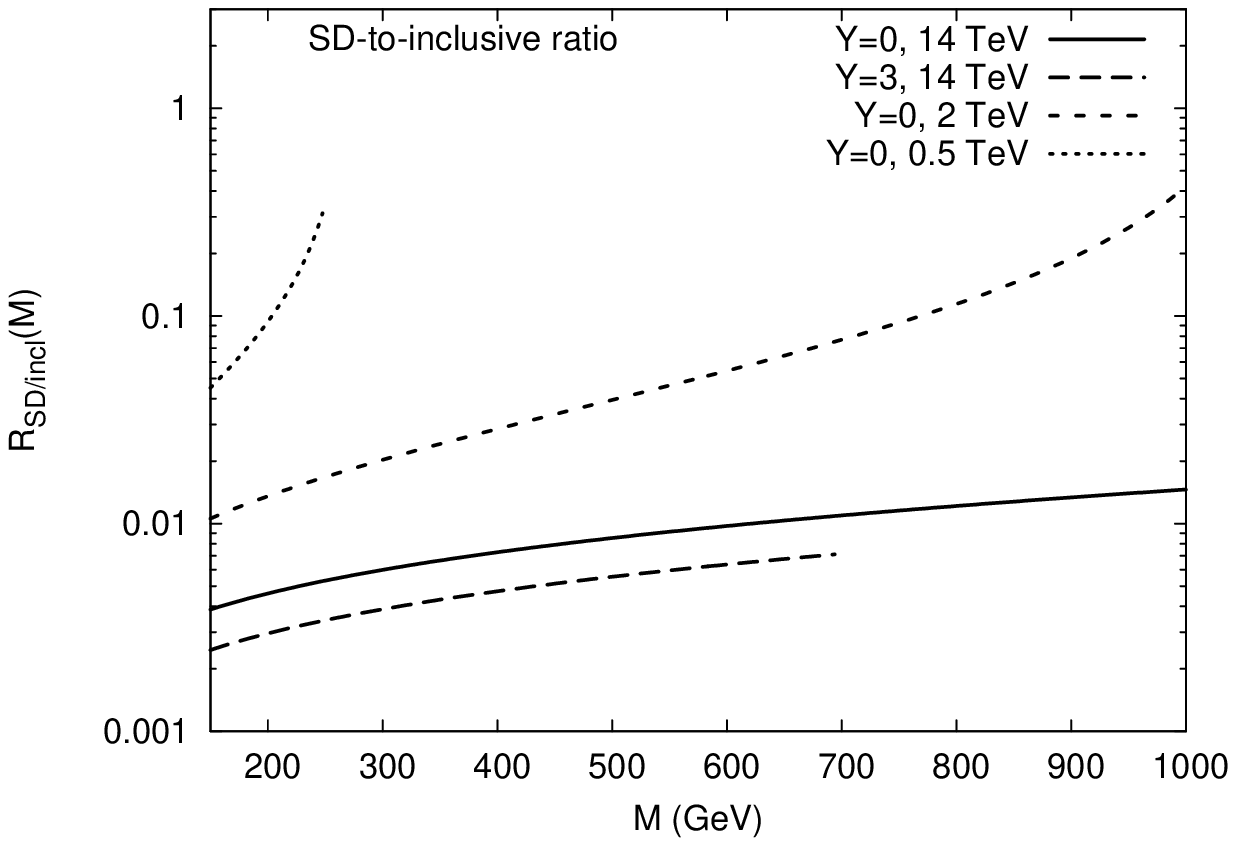}}
\end{minipage}
\begin{minipage}{0.495\textwidth}
 \centerline{\includegraphics[width=1.0\textwidth]{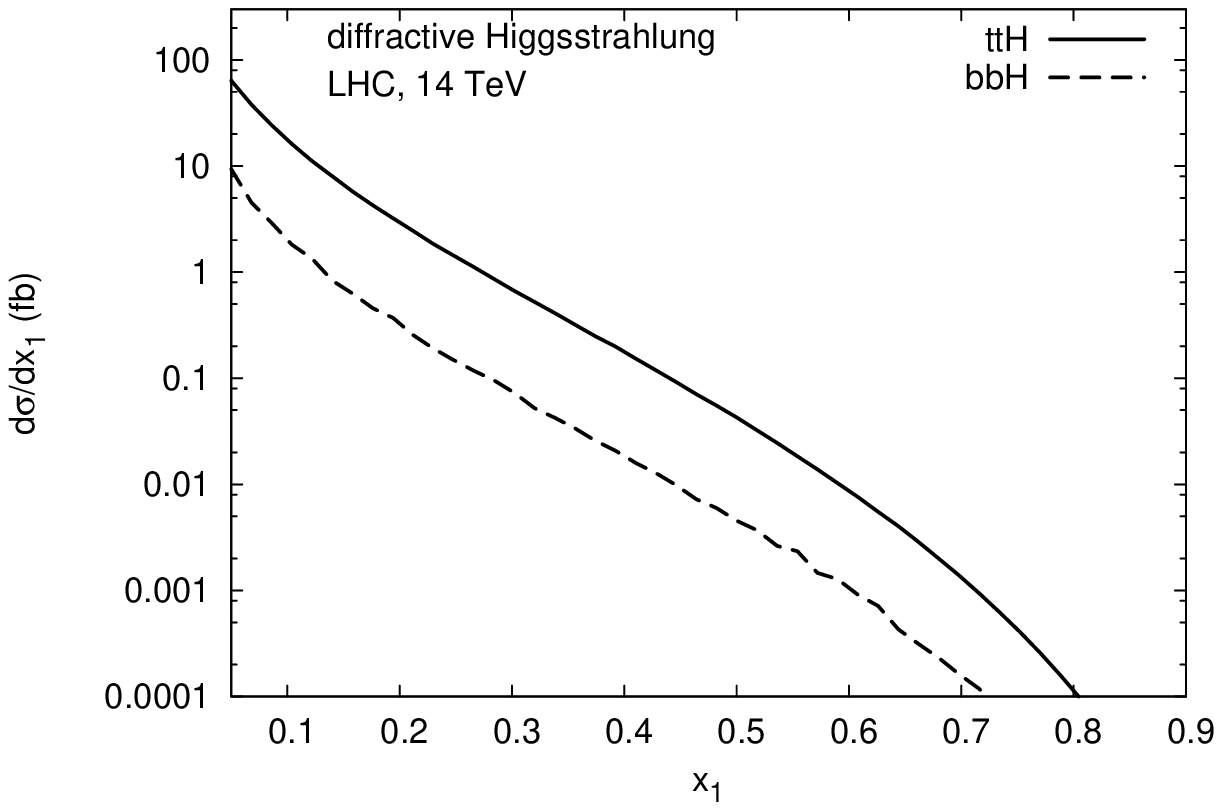}}
\end{minipage}
   \caption{ \small 
  The SD-to-inclusive ratio $R(M)$ as a function of $\bar QQH$ invariant mass $M$ for 
  different c.m. energies $\sqrt{s}=0.5,2,14$ TeV (left panel) and the differential cross 
  section $d\sigma/dx_1$ of the SD Higgsstrahlung off $t\bar{t}$ and $b\bar{b}$ 
  pairs for $\sqrt{s}=14$ TeV (right panel) \cite{Pasechnik:2014lga}.}
 \label{fig:SD-QQh}
\end{figure*}

The hard scales which control the diffractive Higgsstrahlung process are, 
$\langle r^2\rangle \sim 1/m_Q^2$ and $\langle\rho^2\rangle \sim 1/\tau^2$, 
where $\tau^2=M_H^2+\alpha_3 M_{Q\bar Q}^2$ in terms of the Higgs boson mass,
$M_H$, and the $Q\bar Q$ pair invariant mass, $M_{Q\bar Q}$. Another length scale here
is the distance between $i$th and $j$th projectile partons, 
$\langle r_{ij}^2\rangle \sim \langle R\rangle^2$, is soft for light valence/sea quarks in the
proton wave function. Before the hard gluon splits into $\bar QQH$ system 
it undergoes multiple splittings $g\to gg$ populating the projectile 
fragmentation domain with gluon radiation with momenta below 
the hard scale of the process $p_{\perp}^{rad}<M_{\bar QQH}$. 
The latter should be accounted for via a gluon PDF evolution.

The SD-to-inclusive ratio of the cross sections for different c.m. energies 
$\sqrt{s}=0.5,7,14$ TeV and for two distinct rapidities $Y=0$ and $3$ 
as functions of $\bar QQH$ invariant mass $M$ are shown in Fig.~\ref{fig:SD-QQh} (left). 
The ratio is similar to that for heavy quark production \cite{KPST07} and thus 
in good agreement with experimental data from the Tevatron. Note, this ratio has falling energy- 
and rising $M$-dependence, where $M$ is the invariant mass of the produced $\bar QQH$ system. 
This is similar to what was found for diffractive Drell-Yan process 
\cite{Pasechnik:2011nw,Pasechnik:2012ac} and has the same origin, namely, 
breakdown of QCD factorisation and the saturated form of the dipole cross section.

The differential cross sections of single diffractive $b\bar b$ and $t\bar t$ 
production in association with the Higgs boson are shown in Fig.~\ref{fig:SD-QQh} (right) 
as functions of $x_1$ variable at the LHC energy $\sqrt{s}=14$ TeV implied that the Higgs boson 
transverse momentum is large, i.e. $\kappa\gtrsim m_H$. In this case the asymptotic dipole 
formula based upon the collinear projectile gluon PDF (\ref{PDFd}) and the first (quadratic) 
term in the dipole cross section is a good approximation and reproduces the exact 
$k_\perp$-factorisation result for the inclusive Higgsstrahlung transverse momentum 
distribution in both the shape and normalisation (for more details, see Ref.~\cite{Pasechnik:2014lga}). 
The contribution of diffractive gluon excitations to the Higgsstrahlung dominates the total 
Higgsstrahlung cross section due a large yield from central rapidities. Besides, 
the SD Higgsstrahlung off top quarks is larger than that from the bottom 
while shapes of the $x_1$ distributions are similar. Additional radiation of the Higgs 
boson enhances the contribution of heavy quarks and thus 
compensates the smallness of their diffractive production modes. 

Analogically to other diffractive bremsstrahlung processes discussed in previous sections, breakdown of 
QCD factorisation leads to a flatter hard scale dependence of the cross section. This is a result of leading twist 
behaviour which have been discussed above and which has been confirmed by 
the comparison of data on diffractive production of charm and beauty \cite{KPST07}.

\section{Summary}

In this short review, we have discussed major properties and basic dynamics of single diffractive 
processes of $\gamma^*,\,Z^0$ and $W^{\pm}$ bosons production processes at the LHC, 
as well as leading twist heavy flavor hadroproduction at large Feynman $x_F$ and diffractive
Higgsstrahlung off heavy quarks. We outlined the manifestations of diffractive factorisation 
breaking in these single diffractive reactions within the framework of color dipole description,
which is suitable for studies of the interplay between soft and hard fluctuations. The latter reliably
determine diffractive hadroproduction in the projectile fragmentation region.

The first, rather obvious source for violation of diffractive factorisation, is related to the absorptive 
corrections (called sometimes survival probability of large rapidity gaps). The absorptive corrections 
affect differently the diagonal and off-diagonal diffractive amplitudes \cite{KPSdiff,PCAC}, leading to a
breakdown of diffractive QCD factorisation in hard diffractive processes, like diffractive radiation
of heavy Abelian particles and heavy flavors. The dipole approach enables to account for the absorptive
corrections automatically at the amplitude level.

The second, more sophisticated reason for diffractive factorisation breaking, is specific for Abelian 
radiation, namely, a quark cannot radiate in the forward direction (zero momentum transfer),
where diffractive cross sections usually have a maximum. Forward diffraction becomes possible 
due to intrinsic transverse motion of quarks inside the proton, although the magnitude of the 
forward cross section remains very small \cite{KPST06,Pasechnik:2011nw}. A much larger 
contribution to Abelian radiation in the forward direction in $pp$ collisions comes from interaction 
with the spectator partons in the proton. Such a hard-soft interplay is specific for the considered 
processes in variance to the DDIS involving no co-moving spectator partons. 

These mechanisms of diffractive factorisation breaking lead to rather unusual features of 
the leading-twist diffractive Abelian radiation w.r.t. its hard scale and energy dependence.

The outlined sources of factorisation breaking are also presented in diffractive radiation of 
non-Abelian particles. Interactions of the radiated gluon makes it possible to be radiated 
even at zero momentum transfer. These processes have been quantitatively analysed 
in such important channels as diffractive heavy flavor production and Higgsstrahlung in the
projectile fragmentation region. Further studies of these effects, both experimentally and 
theoretically, are of major importance for upcoming LHC measurements.

{\bf Acknowledgments}

This study was partially supported by Fondecyt (Chile) grants 
1120920, 1130543 and 1130549, and by ECOS-Conicyt
grant No. C12E04. R. P. was partially supported by 
Swedish Research Council Grant No. 2013-4287.


\end{document}